\documentclass[longauth, traditabstract]{aa}

\usepackage{graphicx}
\usepackage{natbib}
\bibpunct{(}{)}{;}{a}{}{,} 
\usepackage{txfonts}
\begin{document}
   \title{Planck pre-launch status: calibration of the Low Frequency Instrument flight model radiometers}
  \titlerunning{Calibration of LFI flight model radiometers}

   \author{F.	Villa	\inst{1}	\and
L.	Terenzi	\inst{1}	\and
M.	Sandri	\inst{1}	\and
P.	Meinhold	\inst{2}	\and
T.	Poutanen	\inst{3}\fnmsep\inst{4}\fnmsep\inst{5} \and
P.	Battaglia	\inst{6}	\and
C.	Franceschet	\inst{7}	\and
N.	Hughes	\inst{8}\and
M.	Laaninen	\inst{8}	\and
P.	Lapolla	\inst{6}	\and
M.	Bersanelli	\inst{7}	\and
R.C.	Butler	\inst{1}	\and
F.	Cuttaia	\inst{1}	\and
O.	D'Arcangelo	\inst{9}	\and
M.	Frailis	\inst{10}	\and
E.	Franceschi	\inst{1}	\and
S.	Galeotta	\inst{10}	\and
A.	Gregorio	\inst{11}	\and
R.	Leonardi	\inst{2}	\and
S.R.	Lowe	\inst{12}	\and
N.	Mandolesi	\inst{1}	\and
M.	Maris	\inst{10}	\and
L.	Mendes	\inst{13}	\and
A.	Mennella	\inst{7}	\and
G.	Morgante	\inst{1}	\and
L.	Stringhetti	\inst{1}\fnmsep\thanks{\emph{Present address: } Astrium GmbH, Friedrichshafen, Germany}	\and
M.	Tomasi	\inst{7}	\and
L.	Valenziano	\inst{1}	\and
A.	Zacchei	\inst{10}	\and
A.	Zonca	\inst{14}	\and
B.	Aja	\inst{15}	\and
E.	Artal	\inst{15}	\and
M.	Balasini	\inst{6}	\and
T.	Bernardino	\inst{16}	\and
E.	Blackhurst	\inst{12}	\and
L.	Boschini	\inst{6}	\and
B.	Cappellini	\inst{14}	\and
F.	Cavaliere	\inst{7}	\and
A.	Colin	\inst{16}	\and
F.	Colombo	\inst{6}	\and
R.J.	Davis	\inst{12}	\and
L.	De La Fuente	\inst{15}	\and
J.	Edgeley	\inst{12}	\and
T.	Gaier	\inst{17}	\and
A.	Galtress	\inst{12}	\and
R.	Hoyland	\inst{18}	\and
P.	Jukkala	\inst{8}	\and
D.	Kettle	\inst{12}	\and
V-H. Kilpia	\inst{8}	\and
C.R.	Lawrence	\inst{16}	\and
D.	Lawson	\inst{12}	\and
J.P.	Leahy	\inst{12}	\and
P.	Leutenegger	\inst{6}	\and
S.	Levin	\inst{16}	\and
D.	Maino	\inst{7}	\and
M.	Malaspina	\inst{1}	\and
A.	Mediavilla	\inst{15}	\and
M.	Miccolis	\inst{6}	\and
L.	Pagan	\inst{6}	\and
J.P.	Pascual	\inst{15}	\and
F.	Pasian	\inst{10}	\and
M.	Pecora	\inst{6}	\and
M.	Pospieszalski	\inst{19}	\and
N.	Roddis	\inst{12}	\and
M.J.	Salmon	\inst{16}	\and
M.	Seiffert	\inst{17}	\and
R.	Silvestri	\inst{6}	\and
A.	Simonetto	\inst{9}	\and
P.	Sjoman	\inst{8}	\and
C.	Sozzi	\inst{9}	\and
J.	Tuovinen	\inst{20}	\and
J.	Varis 	\inst{20}	\and
A.	Wilkinson	\inst{12}	\and
F.	Winder	\inst{12}
}
\offprints{F. Villa}

   \institute{INAF - Istituto di Astrofisica Spaziale e Fisica Cosmica, via P. Gobetti, 101, I-40129 Bologna, Italy\\
              \email{villa@iasfbo.inaf.it}
   \and
   Department of Physics, University of California, Santa Barbara, CA 93106-9530, USA
   \and
University of Helsinki, Department of Physics, P. O. Box 64, FIN-00014 Helsinki, Finland.
   \and
Helsinki Institute of Physics, P.\ O.\ Box 64, FIN-00014 Helsinki, Finland.
   \and
Mets\"ahovi Radio Observatory, Helsinki University of Technology, Mets\"ahovintie 114, 02540 Kylm\"al\"a, Finland 
   \and 
   Thales Alenia Space Italia S.p.A., S.S. Padana Superiore 290, 20090 Vimodrone (MI)
     \and
   Universit\`a degli Studi di Milano, Via Celoria 16, I-20133 Milano, Italy
   \and
   DA-Design Oy. (aka Ylinen Electronics), Keskuskatu 29, FI-31600 Jokioinen, Finland
   \and
   IFP-CNR, Via Cozzi 53, I-20125 Milano, Italy 
   \and
   INAF - Osservatorio Astronomico di Trieste, via Tiepolo, 11, I-34143 Trieste, Italy 
   \and
   University of Trieste, Department of Physics - via Valerio, 2 Trieste I-34127, Italy 
   \and
   Jodrell Bank Centre for Astrophysics, Alan Turing Building, The University of Manchester, Manchester, M13 9PL, UK
   \and
   Planck Science Office, European Space Agency, ESAC, P.O. box 78, 28691 Villanueva de la Cañada, Madrid, Spain
\and 
INAF - Istituto di Astrofisica Spaziale e Fisica Cosmica, Via E. Bassini 15, I-20133 Milano, Italy
\and Universidad de Cantabria, Dep. De Ingenieria de Comunicaciones. Av. Los Castros s/n, 39005, Santander-Spain.
\and
   Instituto de Fisica de Cantabria (CSIC-UC) Av. Los Castros s/n 39005 Santander, Spain
  \and
   Jet Propulsion Laboratory,  4800 Oak Grove Drive, Pasadena, California 91109
  \and 
  Instituto de Astrofísica de Canarias C/ Via Lactea, s/n E38205 - La Laguna (Tenerife), Spain    
\and
National Radio Astronomy Observatory, Stone hall, University of Virginia, 520 Edgemont road,  Charlottesville, USA
\and
MilliLab, VTT Technical Research Centre of Finland, Tietotie 3, Otaniemi, Espoo, Finland
}
   \date{Received July 9, 2009; accepted May 3, 2010}

\abstract{The Low Frequency Instrument (LFI) on-board the ESA Planck satellite carries eleven radiometer subsystems, called Radiometer Chain Assemblies (RCAs), each composed of a pair of pseudo-correlation receivers. We describe the on-ground calibration campaign performed to qualify the flight model RCAs and to measure their pre-launch performances.
Each RCA was calibrated in a dedicated flight-like cryogenic environment with the radiometer front-end cooled to 20K and the back-end at 300K, and with an external input load cooled to 4K. A matched load simulating a blackbody at different temperatures was placed in front of the sky horn to derive basic radiometer properties such as noise temperature, gain, and noise performance, e.g.  1/f noise. The spectral response of each detector was measured as was their susceptibility to thermal variation.
All eleven LFI RCAs were calibrated. 
Instrumental parameters measured in these tests, such as noise temperature, bandwidth, radiometer isolation, and linearity, provide essential inputs to the Planck-LFI data analysis.}


\maketitle
%

\section{Introduction}

The Planck mission\footnote{Planck \emph{(http://www.esa.int/Planck)} is a project of the European Space Agency - ESA - with instruments provided by two scientific Consortia funded by ESA member states (in particular the lead countries: France and Italy) with contributions from NASA (USA), and telescope reflectors provided in a collaboration between ESA and a scientific Consortium led and funded by Denmark.} 
has been developed to provide a deep, full-sky image of the cosmic microwave background (CMB) in both temperature and polarization. Planck incorporates an unprecedented combination of sensitivity, angular resolution and spectral range - spanning from centimeter to sub-millimeter wavelengths - by integrating two complementary cryogenic instruments in the focal plane of the Planck telescope. The Low Frequency Instrument (LFI) covers the region below the CMB blackbody peak in three frequency bands centered at 30, 44 and 70 GHz. The spectral range of the LFI is also suitable for a wealth of galactic and extragalactic astrophysics. The LFI maps will address studies of diffuse Galactic free-free and synchrotron emission, emission from spinning dust grains, and  discrete Galactic radio sources. Extragalactic radio sources will also be observed, particularly those with flat or strongly inverted spectra, peaking at mm wavelengths. Furthermore, the Planck scanning strategy will also allow monitoring of radio source variability on a variety of time scales. While these are highly interesting astrophysical objectives, the LFI design and calibration is driven by the main Planck scientific scope, i.e., CMB science. 

Low Frequency Instrument is an array of cryogenic radiometers based on indium phospide (InP) cryogenic HEMT low noise amplifiers \citep{bersanelli2009}. The array is composed of 22 pseudo--correlation radiometers mounted in eleven independent radiometer units called "radiometer chain assemblies" (RCAs), two centered at 30 GHz, three at 44 GHz and six at 70 GHz 
\footnote{The chains are numbered as {\tt RCAXX} where {\tt XX} is a number from {\tt 18} to {\tt 23} for the 70 GHz RCAs, from {\tt 24} to {\tt 26} for the RCAs at 44 GHz, and from {\tt 27} to {\tt 28} for the RCAs at 30 GHz}. To optimize the sensitivity and minimize the power dissipation in the front end, each RCA is split into a front-end module (FEM), cooled to 20K, and a back-end module (BEM), operating at 300K, connected by a set of composite waveguides.

Accurate calibration is mandatory for optimal operation of the instrument during the full-sky survey and for measuring parameters that are essential for the Planck data analysis.

The LFI calibration strategy has been  based on a complementary approach that includes both pre-launch and post-launch activities. On-ground measurements were performed at all-unit and sub-unit levels, both for qualification and performance verification. Each single FEM and BEM as well as the passive components (feed horn, orthomode transducers, waveguides, 4K reference loads) were tested in a stand--alone 
configuration before they were integrated into the RCA units.

The final scientific calibration of the LFI was carried out at two different integration levels, depending on the measured parameter. First each RCA was tested independently in dedicated cryofacilities, which are capable of reaching a temperature of $\sim$4K at the external input loads. These conditions were necessary for an accurate measurement of key parameters such as system noise temperature, bandwidth, radiometer isolation, and linearity. Subsequently, the eleven RCAs were integrated into the full LFI instrument (the so-called Òradiometer array assemblyÓ, RAA) and tested as a complete instrument system in a large cryofacility, with highly stable input loads cooled down to 20K.

We reports on the calibration campaign of the RCAs, while \citet{mennella2009a} report on the RAA calibration. The parameters derived here are crucial for the Planck-LFI scientific analysis. Noise temperatures, bandwidths, and radiometer isolation provide essential information to construct an adequate noise model, which is needed as an input to the map-making process. Any non-linearity of the instrument response must be accurately measured, because corrections may be needed in the data analysis, particularly for observations of strong sources such as planets (crucial for in-flight beam reconstruction) and the Galactic plane. As part of the RCA testing, we also performed an end-to-end measurement of the bandshape inside the cryofacility. Finally, for comparison and as a consistency check, the RCA test plan also included measurement of parameters whose primary calibration relies on the RAA test campaign, such as optimal radiometer bias (tuning), 1/f noise (knee frequency and slope), gain, and thermal susceptibility.

The 30 and 44 GHz RCAs were integrated and tested in Thales Alenia Space Italia (TAS-I), formerly Laben, from the beginning of January 2006 to end of May 2006 using a dedicated cryofacility \citep{terenzi2009a} to reproduce flight-like thermal interfaces and input loads. The 70 GHz RCAs were integrated and calibrated in Yilinen Electronics (Finland) from the end of April 2005 to mid February 2006 with a similar cryofacility, but with simplified thermal interfaces \citep{terenzi2009a}. The differences between the two cryofacilities resulted in slightly different test procedures, because the temperatures were not controlled in the same way. 

Section \ref{Concept} of the paper describes the concept of RCA calibration, while Sect. \ref{cryo} illustrates the two cryofacilities. 
In Sect. \ref{results} we describe each test, the methods used in the analysis and the results. The conclusions are given in Sect. \ref{conclusions}.

\section{Main concepts and calibration logic} \label{Concept} 
\subsection{Radiometer chain assembly description}

\begin{figure*}
\centering
 \includegraphics[width=\textwidth]{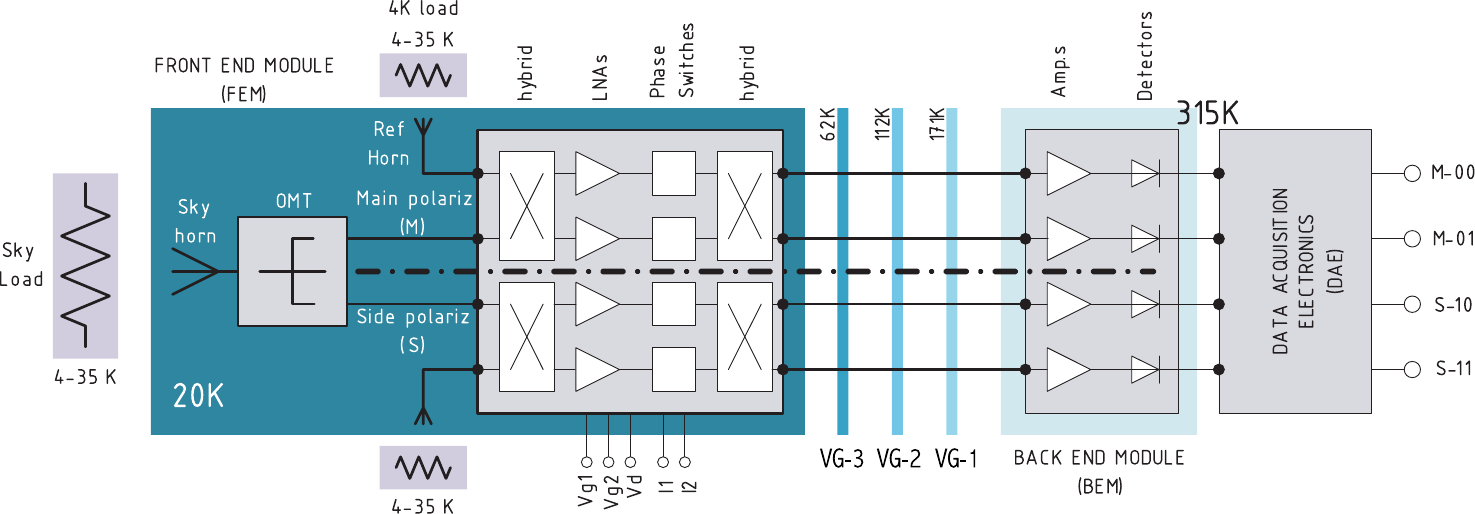}
   \caption {Scheme of the RCA and its flight-like thermal interfaces. The FEM is at 20K, while the BEM is at 315K. At 30 and 44 GHz the thermal interface attached to the third and coldest V-groove (VG-3) is controlled at a temperature near 60K, while for the 70 GHz RCA the VG-3 interface was not controlled in temperature. Both loads - the sky load and the reference load - are controlled in a temperature of approximately 4K to 35K.}
   \label{fig:rca_scheme}
\end{figure*}

A diagram of an RCA is shown in Fig. \ref{fig:rca_scheme}.
A corrugated feed horn \citep{villa2009}, which collects the radiation from the telescope, $T_{sky}$, is connected to an ortho-mode transducer \citep{darcangelo2009a}, which divides the signal into two orthogonal polarizations, namely 'M' (main) and 'S'  (side) branches. The OMT is connected to the front-end section \citep{davis2009, varis2009}, in which each polarization of the sky signal is mixed with the signal from a stable reference load, $T_{ref}$, via a hybrid coupler \citep{valenziano2009}. The signal is then amplified by a factor $G_{fe}$ and shifted in phase by 0--180 degrees at 8kHz synchronously with the acquisition electronics. Finally, a second hybrid coupler separates the input sky signal from the reference load signal.

1.75 meter long waveguides \citep{darcangelo2009b} are connected to the FEM in bundles of four elements providing the thermal break between 20K and 300 K where the ambient temperature back-end section of the radiometer is located \citep{artal2009, varis2009}. The waveguides are thermally attached to the three thermal shields of the satellite, the V-grooves. They act as radiators to passively cool down the payload to about 50K. They drive the thermal gradient along the waveguides.

Inside the BEM the signal is further amplified by $G_{be}$ and then detected by four output detector diodes\footnote{According to the name convention, diodes refer to the Main-polarization are labeled as {\tt M-00, M-01} and those referring to the Side-polarization are labeled as {\tt S-10, S-11}.}. 
In nominal conditions, each of the four diodes detects a voltage alternatively (each $122\mu$sec which corresponds to $1/8192$ Hz$^{-1}$) proportional by a factor $a$ to the sky load and reference load temperature. By differencing these two signals a very stable output is obtained, which allows the measurement of very faint signals. 

Assuming negligible mismatches between the two radiometer legs and within the phase switch, the differenced radiometer output at each detector averaged over the bandwidth $\beta$ can be written in terms of overall gain,  $G_{tot}$ in units of V/K, and noise temperature,  $T_{N}$, as 

\begin{eqnarray}
V_{out} = &G_{tot}& \left [ \left ( \widetilde{T}_{sky} + T_{N}\right ) - r \cdot
\left ( \widetilde{T}_{ref} + T_{N}\right ) \right ] \label{eq:basic}\\  
&G_{tot} &= a \cdot k \cdot \beta \cdot G_{fe} L_{wg} G_{be} \nonumber \\ 
&T_{N}&\simeq T^{(fe)}_{N} + {T^{(be)}_{N} \over G_{fe}} . \nonumber
\end{eqnarray}
\noindent  
Here we further define the waveguide losses as $L_{wg}$, the front-end noise temperature a $T^{(fe)}_{N}$, and the back-end noise temperature as $T^{(be)}_{N}$, and with $k$, the Boltzmann constant. The noise  contribution of the waveguides due to its attenuation is negligible and not considered here. 

The ohmic losses of the feedhorn -- OMT assembly,  $L_{fo}$, and of the 4K Reference load system, $L_{4K}$, modify the actual sky and reference load through the following equations 
\begin{eqnarray}
\widetilde{T}_{sky} = {T_{sky} \over L_{fo}} + \left ( 1 - {1\over L_{fo}}\right ) T_{phys} \label{eq:skyT}\\
\widetilde{T}_{ref} = {T_{ref} \over L_{4K}} + \left ( 1 - {1\over L_{4K}}\right ) T_{phys}, \label{eq:refT}
\end{eqnarray}
\noindent where $T_{phys}$ its the physical temperature (close to $20$K at operational conditions). 

The $r$ factor in Eq. (\ref{eq:basic}) is the gain modulation factor calculated as 
\begin{equation}
r = {\widetilde{T}_{sky} + T_{N} \over \widetilde{T}_{ref} + T_{N}},
\label{eq:r}
\end{equation}
which nulls the radiometer output. 
A more general form of the averaged power output, which takes into account various non-ideal behaviors of the radiometer components, can be found in  \citet{seiffert2002} and \citet{mennella2003}. The key parameters needed to reconstruct the required signal (differences of $T_{sky}$ from one point of the sky to another) are therefore the photometric calibration, $G_{tot}$, and the gain modulation factor, $r$, which is used to suppress the effect of 1/f noise. Deviations from this first approximation are treated as systematic effects. 

\subsection{Signal model}
To better understand the purpose of the calibration, it is useful to write the Eqs. (\ref{eq:basic}) to (\ref{eq:r}) appropriately so that the attenuation coefficients are taken into account in the RCA parameters instead of considering their effects as a target effective temperature. For the sky signal the output can be written as

\begin{eqnarray}
V^{sky}_{out} &=& \widetilde{G}^{sky}_{tot} \left ( T_{sky} + \widetilde{T}^{sky}_N \right ), \label{eq:sky1}\\
\widetilde{G}^{sky}_{tot} &=& a \cdot k \cdot \beta {1\over L_{fo}} G_{fe} {1\over L_{wg}} G_{be}, \\
\widetilde{T}^{sky}_{N}  &=& T_{N} + \left [ \left (L_{fo}-1\right )T_{phys}\right ],
\end{eqnarray}
and equivalently for the reference signal
\begin{eqnarray}
V^{ref}_{out} &=& \widetilde{G}^{ref}_{tot} \left ( T_{ref} + \widetilde{T}^{ref}_N \right ), \label{eq:ref1}\\
\widetilde{G}^{ref}_{tot} &=& a \cdot k \cdot \beta {1\over L_{4K}} G_{fe} {1\over L_{wg}} G_{be}, \\
\widetilde{T}^{ref}_{N}  &=& T_{N} + \left [ \left (L_{4K}-1\right ) T_{phys}\right ].
\end{eqnarray}

Differencing Eqs.  (\ref{eq:sky1}) and (\ref{eq:ref1}) we obtain the differenced (sky - ref) output similar to Eq. (\ref{eq:basic})

\begin{eqnarray}
V_{out} &=& G^{*}_{tot} \left [ \left ( T_{sky} + \widetilde{T}^{sky}_N \right ) - r^* 
\left ( T_{ref} + \widetilde{T}^{ref}_N \right )\right ],   \label{eq:diff1} \\
G^*_{tot} &=& G_{tot} {1 \over L_{fo}}, \\
r^*       &=& r {L_{fo} \over L_{4K}}.
\end{eqnarray}

Equations  (\ref{eq:sky1}), (\ref{eq:ref1}), and (\ref{eq:diff1}) are the basis for the RCA calibration, because all parameters involved were measured during the RCA test campaign by stimulating each RCA at cryo temperature (close to the operational in-flight conditions) with several known $T_{sky}$ and $T_{ref}$ values. 

\subsection{Radiometer chain assembly calibration plan}
Each RCA calibration included (i) functional tests, to verify the functionality of the RCA; (ii) bias tuning, to set the best amplifier gains and phase switch bias currents for maximum performance (i.e. minimum noise temperature and best radiometer balancing); (iii) basic radiometer property measurements to estimate $G^*$, $\widetilde{T}^{sky}_N$, $\widetilde{T}^{ref}_N$; (iv) noise performance measurements to evaluate $1/f$, white noise level  and the $r^*$ parameter; (v) spectral response measurements to derive the relative bandshape; (vi) susceptibility measurements of radiometer  thermal variations to estimate the dependence of noise and gain with temperature. The list of tests is given in Table \ref{tab:logic} together with a brief description of the purpose of each test.  

\begin{table}[h]
\begin{minipage}[t]{\columnwidth}
\caption{Calibration test list. The first column reports the test identification. In the second column the purpose of each test is described; WN is the white noise level; $f_{k}$, $\alpha$ are the 1/f knee frequency and slope respectively; 
$\beta$ is the equivalent radiometer bandwidth derived from noise; $r$ is the modulation factor. Apart from the first test {\tt RCA\_AMB}, which is performed at ambient temperature, all the other tests are performed at the operational temperature (i.e. at a temperature as close as possible to in-orbit conditions).  
} 
\label{tab:logic}      
\centering                          
\renewcommand{\footnoterule}{}  
\begin{tabular}{l l}        
\hline\hline                       
{\sc Test id} & {\sc Description} \\
\hline
\multicolumn{2}{c}{\sc Functionality} \\
{\tt RCA\_AMB} & Functional test at ambient \\
{\tt RCA\_CRY} & Functional test at cryo \\
\hline
\multicolumn{2}{c}{\sc Tuning} \\
{\tt RCA\_TUN} & Gain and offset tuning of the DAE \\
               & Tuning of the front end module \\
               & (phase switches and gate voltages) \\
\hline
\multicolumn{2}{c}{\sc Basic Properties} \\
{\tt RCA\_OFT} & Radiometer offset \\
               & measurement  \\
{\tt RCA\_TNG} & Noise temperature and \\
               & photometric gain \\
{\tt RCA\_LIS} & Radiometer linearity \\
\hline
\multicolumn{2}{c}{\sc Noise Properties} \\

{\tt RCA\_STn} & Noise performances tests: \\
                & WN, $f_k$ and $\alpha$, $\beta$, $r$ \\
{\tt RCA\_UNC} & Verification of the effect  \\
               & of the radiometer switching \\
               & on the noise spectrum \\
\hline
\multicolumn{2}{c}{\sc Band Pass Response} \\
{\tt RCA\_SPR} & Bandpass \\
\hline 
\multicolumn{2}{c}{\sc Susceptibility} \\
{\tt RCA\_THF} & Susceptibility to \\
               & FEM temperature variations \\
{\tt RCA\_THB} & Susceptibility to  \\
               & BEM Temperature variations \\
{\tt RCA\_THV} & Susceptibility to V-groove \\
               & temperature variations \\
\hline                                  
\end{tabular}
\end{minipage}
\end{table}

Although the calibration plan was the same for all RCAs, the differences in the setup between 30/44 GHz and 70 GHz resulted in a different test sequence and procedures, which retained the  objective of RCA calibration unchanged.

At 70 GHz the RCA tests were carried out in a dedicated cryogenic chamber developed by DA-Design\footnote{\tt http://www.da-design.fi/space} (formerly Ylinen Electronics), which was capable of accommodating two RCAs at one time. Thus the RCA test campaign was planned for three RCA pairs, namely {\tt RCA18} and {\tt RCA23}, {\tt RCA19} and {\tt RCA20}, {\tt RCA21} and {\tt RCA22}. At 30 and 44 GHz, only one RCA at a time was  calibrated. Due to the different length of the waveguides, the thermal interfaces were not exactly the same for different RCAs, which resulted in a slightly different thermal behavior of the cryogenic chamber and thus slightly different calibration conditions.

\section{Radiometer chain assembly calibration facilities}\label{cryo}
In both cases, the calibration facilities used the cryogenic chamber described in \citet{terenzi2009a}, a calibration load, the skyload \citep{terenzi2009b}, electronic ground support equipment (EGSE) and software \citep{malaspina2009}. The heart of the EGSE was  a breadboard of the LFI flight Data Acquisition Electronics (DAE) with Labwindow\texttrademark\footnote{{\tt http://www.ni.com/lwcvi/}} software to control the power supplies to the FEMs and BEMs, read all the housekeeping parameters and digitize the scientific signal at 8KHz without any average or time integration.
The EGSE sent data continuously to a workstation operating the Rachel (Radiometer Chain Evaluator) software for quick-look analysis and data storage \citep{malaspina2009}. 
The data files were stored in FITS format. As two chains were calibrated at the same time at 70 GHz, separate EGSEs and analysis workstations were used for each RCA.  
Below the cryofacilities and skyloads are summarized with the emphasis  on the issues related to the analysis of the calibration data. 
 
\subsection{The cryofacility for the 30 and 44 GHz RCAs}\label{facility_3044}
The chamber with its overall dimensions of $2.0 \times 1.2 \times 1.0$ m$^3$ was able to accept one RCA at a time. 
The chamber was designed to allow  the pressure to reach less than $10^{-5}$ mBar, and contained seven thermal interfaces to reproduce the flight-like thermal conditions of an RCA. During tests it was possible to control and stabilize the BEM temperature, the waveguide-to-spacecraft interface temperature, and the FEM temperature. In addition the two reference targets (the reference load and the sky load) were controlled in temperatures in the range $4-35$K to allow temperature stepping for radiometer linearity tests ({\tt RCA\_LIS}). 
In addition to  the electrical connections for the DAE breadboard  and to control the thermal interfaces, two thermal-vacuum feedthroughs (one for the 30 GHz and the other one for the 44 GHz RCAs) with Kapton windows were provided to  allow access  for the RF signal for the bandpass tests ({\tt RCA\_SPR}).

\begin{figure}
\centering
 \includegraphics[width=8.7cm]{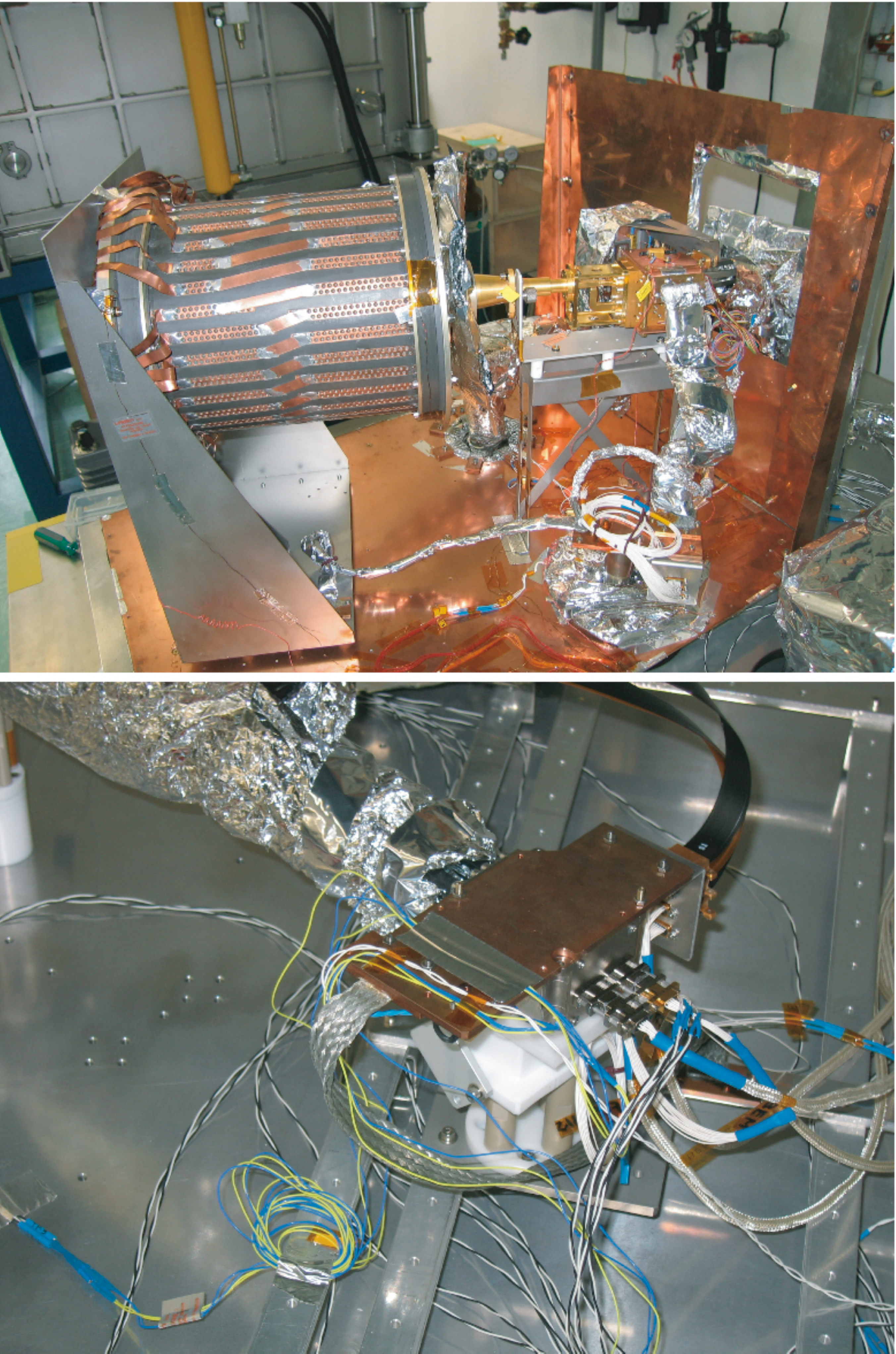}
   \caption{Radiometer chain assembly integrated into the 30 and 44 GHz  cryofacility for calibration. In the picture at the top the skyload facing the horn is visible together with the FEM insulated from the 50K shroud (the copper box). In the bottom picture the BEM and its thermal interface are shown. See the text for details of the cryochamber. 
}
   \label{fig:cryo_3044}
\end{figure}

During the {\tt RCA27} and {\tt RCA28} calibrations an uncertainty in the reference targets' temperature was experienced. 
A visual inspection of the cryochamber after the {\tt RCA28} test gave a possible
explanation and in the {\tt RCA27} test, an additional sensor was put on the back of one of the reference targets in order to verify the probable source of the problem.
The observed behavior was consistent with an excess heat flow through the 4K reference load (4KRL), via its insulated support caused 
by a contact created during cooldown.  A dedicated thermal model was thus developed to derive the Eccosorb  4KRL temperature, $T_{ref}$ from the back plate controller sensor temperature,  $T^{ctrl}_{ref}$ \citep{terenzi2009a}. A quadratic fit was found with $T_{ref} = a+b \cdot T_{ref}^{ctrl} + c\cdot \left ( T^{ctrl}_{ref}\right )^2$ for each pair of detectors coupled to the same radiometer arm and for each 30 GHz RCA. The coefficients derived from the fit are shown in Table \ref{tab:Rfit}.

\begin{table}
\begin{minipage}[t]{\columnwidth}
\caption{Reference load target temperature. Quadratic fit coefficients} 
\label{tab:Rfit}      
\centering                          
\renewcommand{\footnoterule}{}  
\begin{tabular}{l c c}        
\hline\hline                       
   & \multicolumn{2}{c}{\tt RCA27} \\
   & M & S \\   
\hline
a	& $5.7\pm	0.2$ & $2.5\pm	0.1$  \\
b	& $0.58\pm	0.02$ & $0.81\pm	0.01$  \\
c	& $0.00686\pm 5.8\cdot10^{-4}$ & $0.00322\pm	3.2\cdot10^{-4}$ \\
\hline                                  
   &\multicolumn{2}{c}{\tt RCA28} \\
  & M & S \\   
\hline
a	& $2.47\pm	0.05$ & $5.50\pm	0.09$ \\
b	& $0.799\pm	0.007$ & $0.57\pm	0.01$ \\
c &	$0.00407\pm 2.5\cdot10^{-4}$ & $0.00831\pm 4.1\cdot10^{-4}$ \\
\hline
\end{tabular}
\end{minipage}
\end{table}

\subsection{Sky load at 30 and 44 GHz}
\begin{figure}
\centering
 \includegraphics[width=8.5cm]{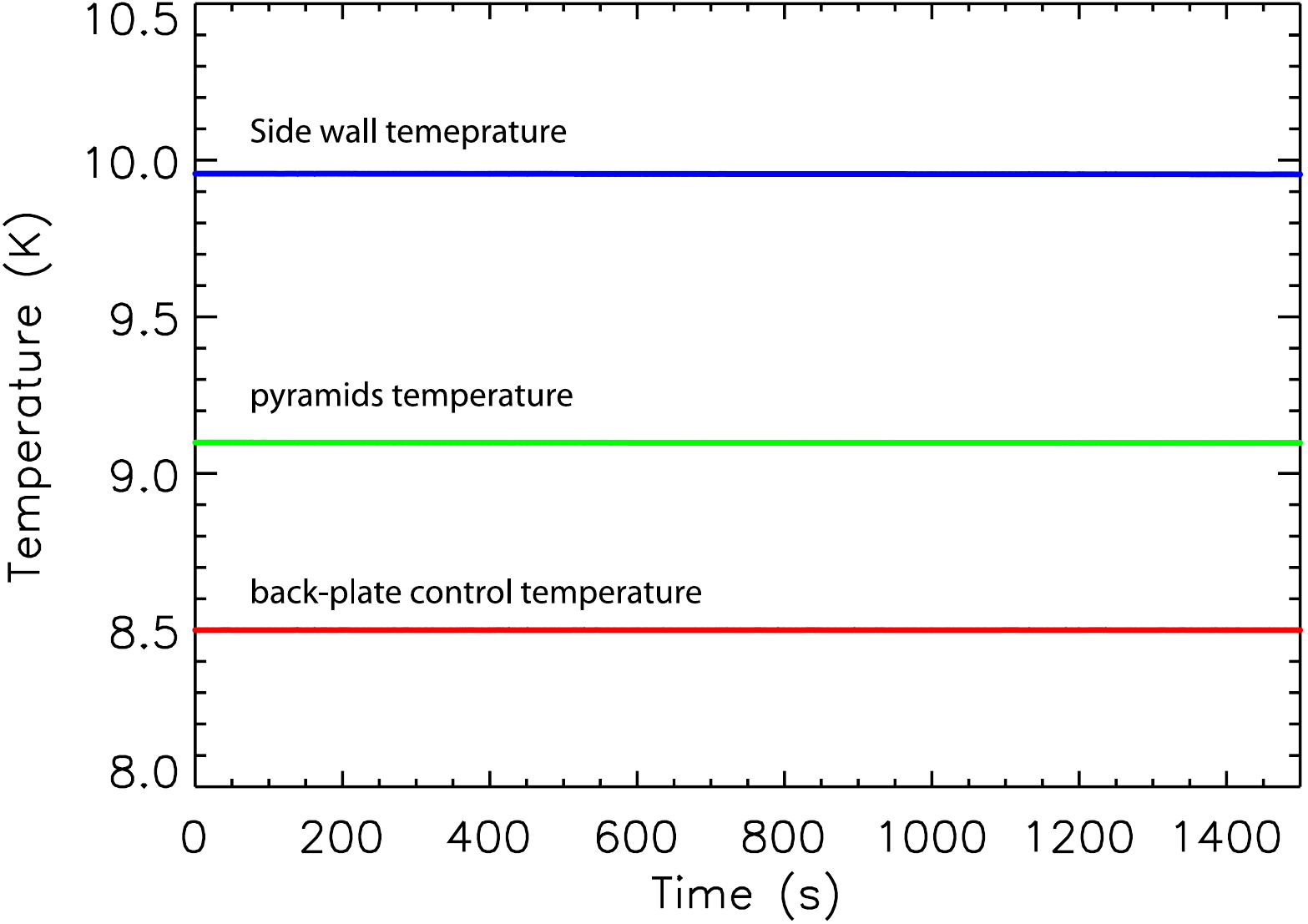}
   \caption{Thirty minutes of data acquired during $\tt RCA28$. Noise temperature and linearity tests are shown with $T_{sky}^{ctrl}$ in red, $T_{sky}$ in green, and $T_{sky}^{side}$ in blue. This represents the worst case of these differences. The stability of the temperatures with the values    
 $T_{sky}^{ctrl}=8.50000\pm 0.00006$,  $T_{sky}=9.0981\pm 0.0004$, and $T_{sky}^{side}=9.9560\pm0.0007$ are also evident.}
   \label{fig:Tpyr-Tside}
\end{figure}

\begin{figure}
\centering
 \includegraphics[width=8.7cm]{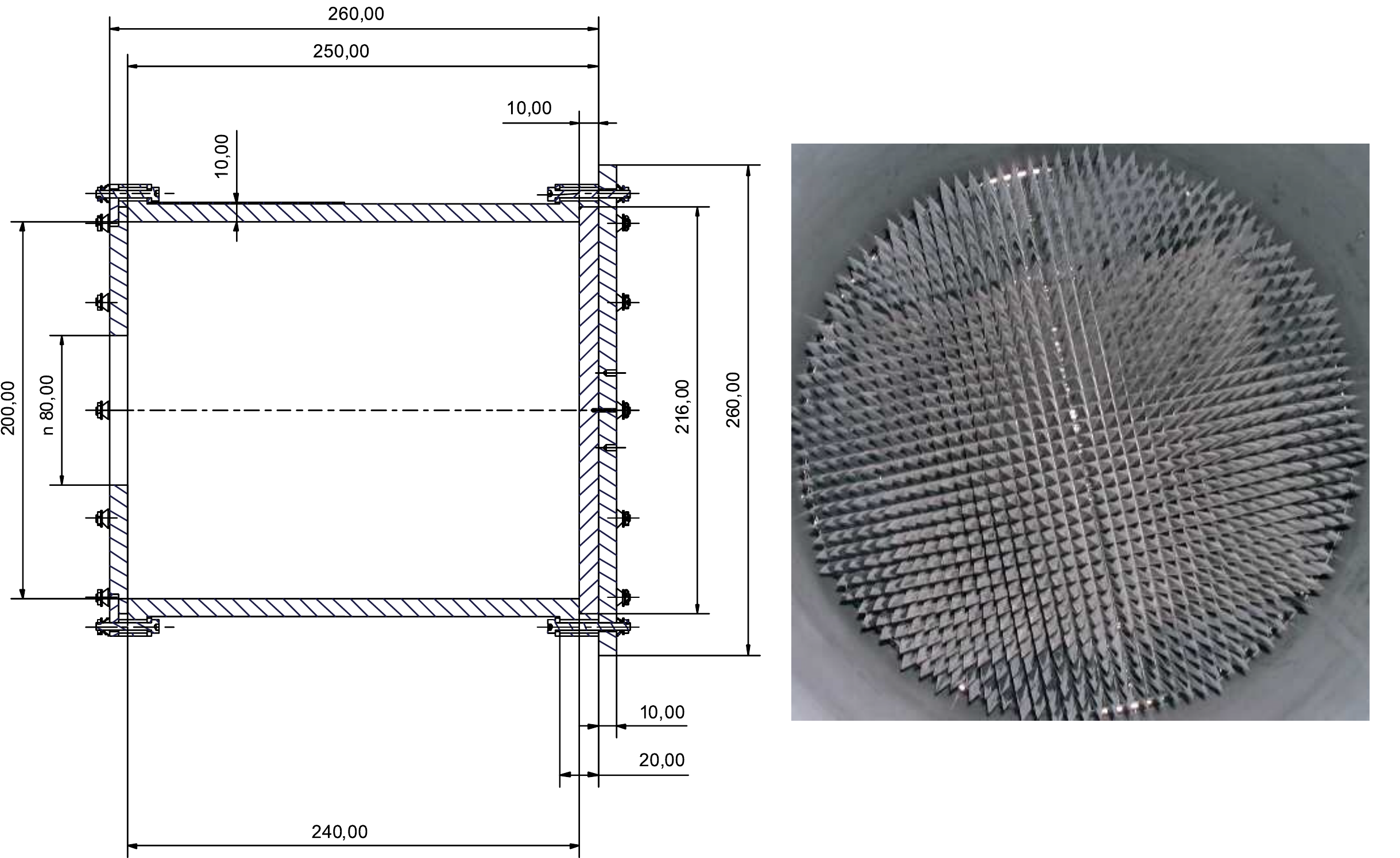} 
   \caption{Bologna design of the RCA sky load calibrator. The overall dimensions in mm are reported in the drawing on the left. The pyramids on the bottom of the skyload are clearly visible on the right picture. 
   }
   \label{fig:bologna_sky_load}
\end{figure}

The calibrator consisted of a cylindrical cavity with walls covered in Eccosorb CR110\footnote{Emerson \& Cuming, www.eccosorb.com} (see Fig. \ref{fig:bologna_sky_load}) . 
The back face of the cavity was covered with Eccosorb pyramids to guarantee a return loss of about $-30$dB.
Details of the skyload are reported in \citet{cuttaia2005}. Four  temperature sensors were placed on the sky load, but only two cernox sensors were taken as reference for calibration.  The first one 
was placed on the back plate of the sky load to measure the temperature of the PID control loop of the sky load, $T^{ctrl}_{sky}$. The second was placed on the Eccosorb pyramids inside the black body cavity and was assumed as the blackbody reference temperature, $T_{sky}$. The contribution to the effective emissivity due to the pyramids was estimated by \citet{cuttaia2005} to be $0.9956$ for the 30GHz channel and $0.9979$ for the 44GHz channel. The effective emissivity was calculated assuming the horn near field pattern and the emissivity of the material. In the case of the skyload side walls the effective emissivity is $4.29\cdot 10^{-3}$ and $2.07\cdot 10^{-3}$ for the 30GHz and 44GHz respectively. In the data analysis only the contribution of the pyramids was considered assuming its emissivity equal to $1$.  Assuming the emissivities reported above and the temperatures as in Fig. \ref{fig:Tpyr-Tside}, the approximation leads to an uncertainty in the brightness temperature of about $0.04$K and the same uncertainty in the noise temperature measurements.

\begin{figure}
\centering
 \includegraphics[width=8.5cm]{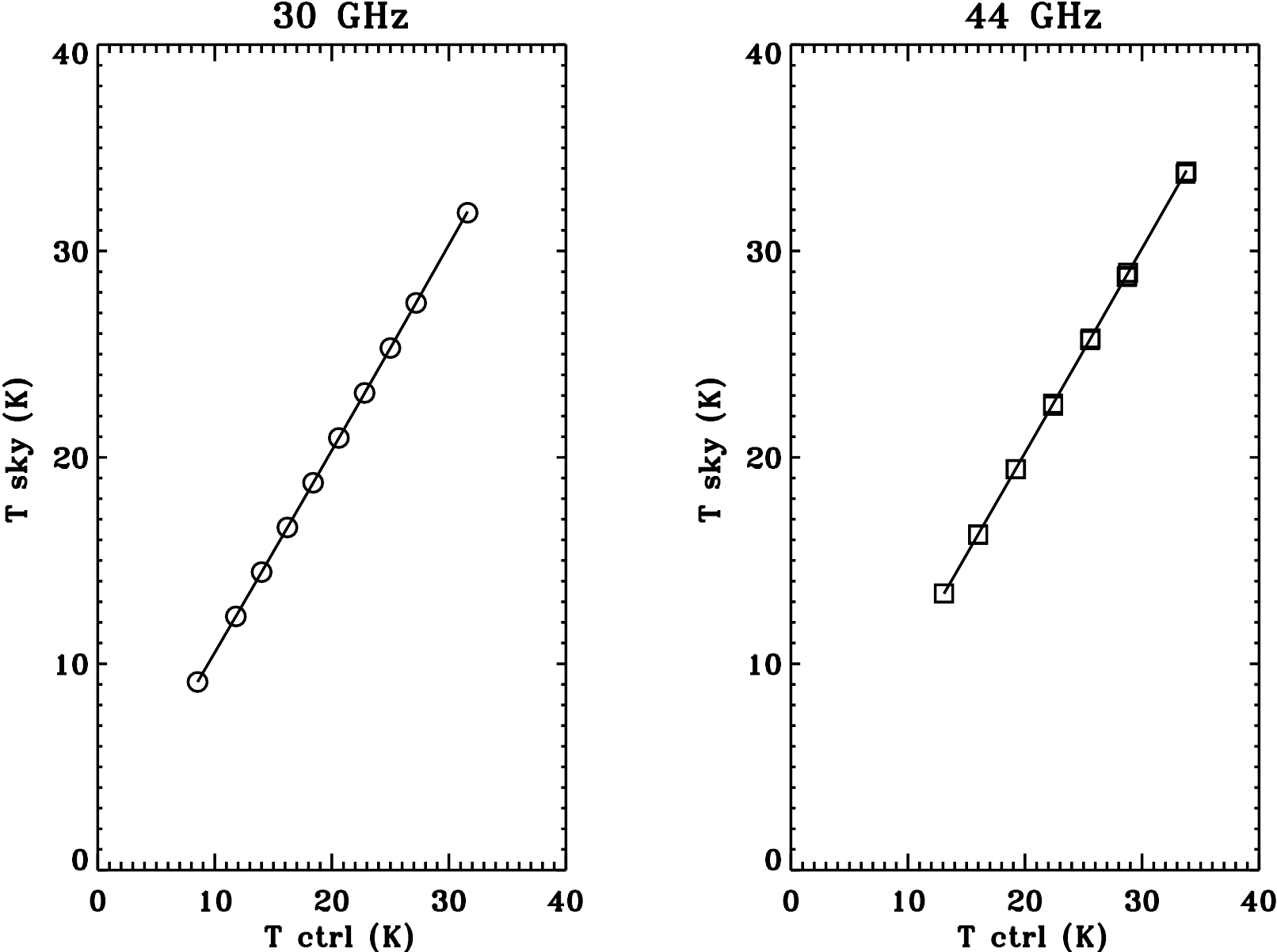}
   \caption{$T_{sky}$ as a function of back plate controller skyload temperature $T^{ctrl}_{sky}$. The left plot refers to the $30$GHz RCAs, based on {\tt RCA28} data (circles). The right plot refers to the $44$GHz RCAs, based on {\tt RCA25} and {\tt RCA26} data (squares). The lines are the quadratic fit to the data.  
   }
   \label{fig:Tctrl-Tsky}
\end{figure}

\begin{figure}
\centering
\includegraphics[width=8.7cm]{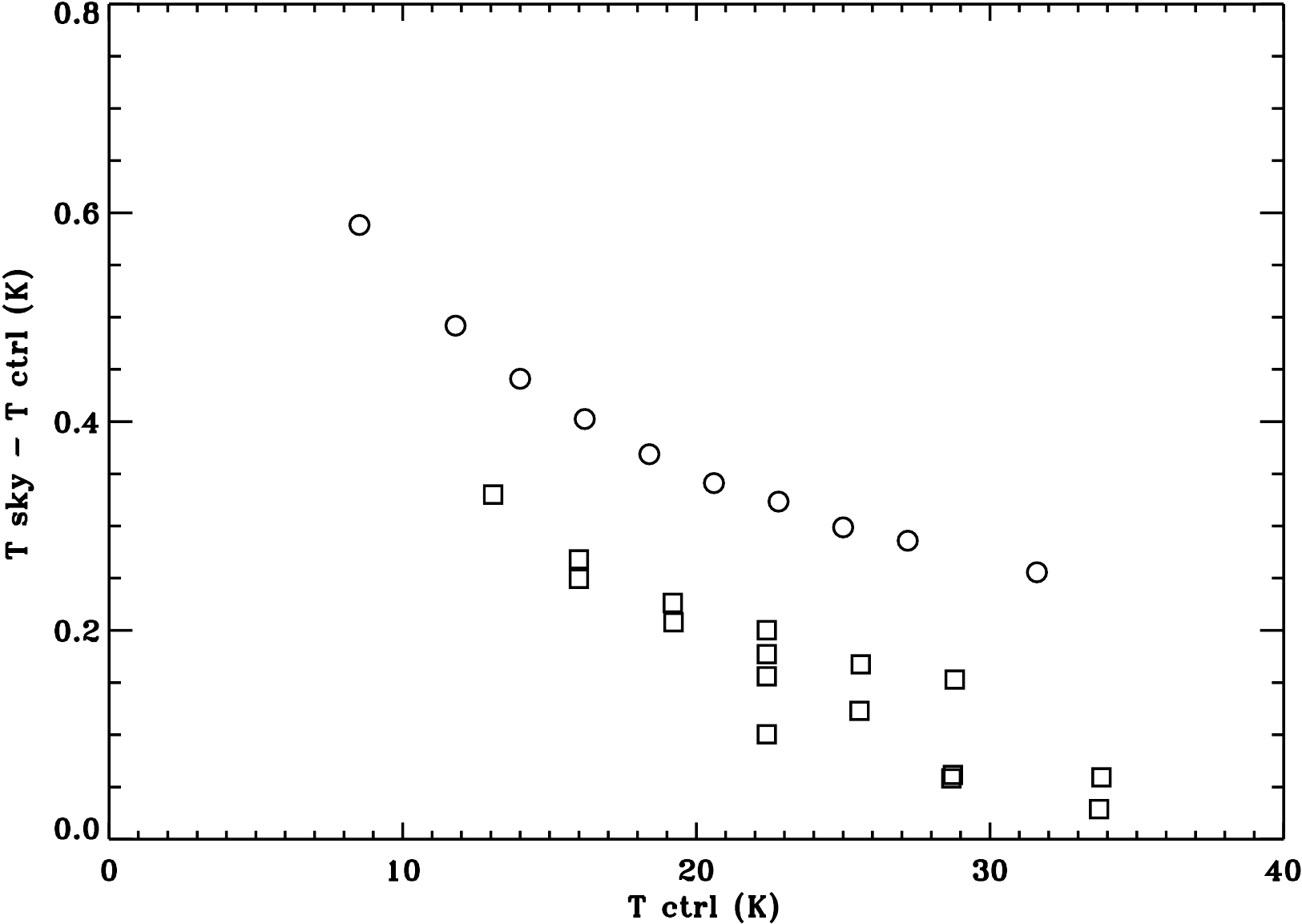} 
   \caption{Differences between $T_{sky}$ and $T^{ctrl}_{sky}$ as a function of $T^{ctrl}_{sky}$ showing the non-linear behavior of  the difference. Circles are for $30$GHz RCAs and squares for $44$GHz RCAs.
   }
   \label{fig:Tdiff}
\end{figure}

Due to a failure in the sensor on the pyramids an analytical evaluation of $T_{sky}$ temperature from $T^{ctrl}_{sky}$ temperature was performed during the calibration of {\tt RCA24} and {\tt RCA27}. The data are shown in Fig. \ref{fig:Tctrl-Tsky}. Although the data show a linear behavior, the differences between $T^{ctrl}_{sky}$ and $T_{sky}$ decrease as the temperature increases (see Fig. \ref{fig:Tdiff}) as expected from the thermal behavior of the system, suggesting that a quadratic fit with $T_{sky} = a+b \cdot T^{ctrl}_{sky} +c\cdot \left ( T^{ctrl}_{sky}\right )^2$ is more representative. This quadratic fit was performed, and the coefficients are reported in Table \ref{tab:Tfit}. 

\begin{table}
\begin{minipage}[t]{\columnwidth}
\caption{Temperarture of the sky load pyramids. Quadratic fit coefficients}  
\label{tab:Tfit}      
\centering                          
\renewcommand{\footnoterule}{}  
\begin{tabular}{l c c}        
\hline\hline                 
   & $30$GHz & $44$GHz \\
a  &$0.9185 \pm 0.0006$&$0.5430 \pm 0.008$ \\
b  &$0.9540 \pm 0.0001$&$0.9795 \pm 0.0008$ \\
c  &$0.0008460 \pm 4.4\cdot 10^{-6}$&$0.000217\pm 1.9\cdot 10^{-5}$\\

\hline                                   
\end{tabular}
\end{minipage}
\end{table}

\subsection{The cryofacility of the 70 GHz RCAs}\label{facility_70}

\begin{figure}
\centering
 \includegraphics[width=8.7cm]{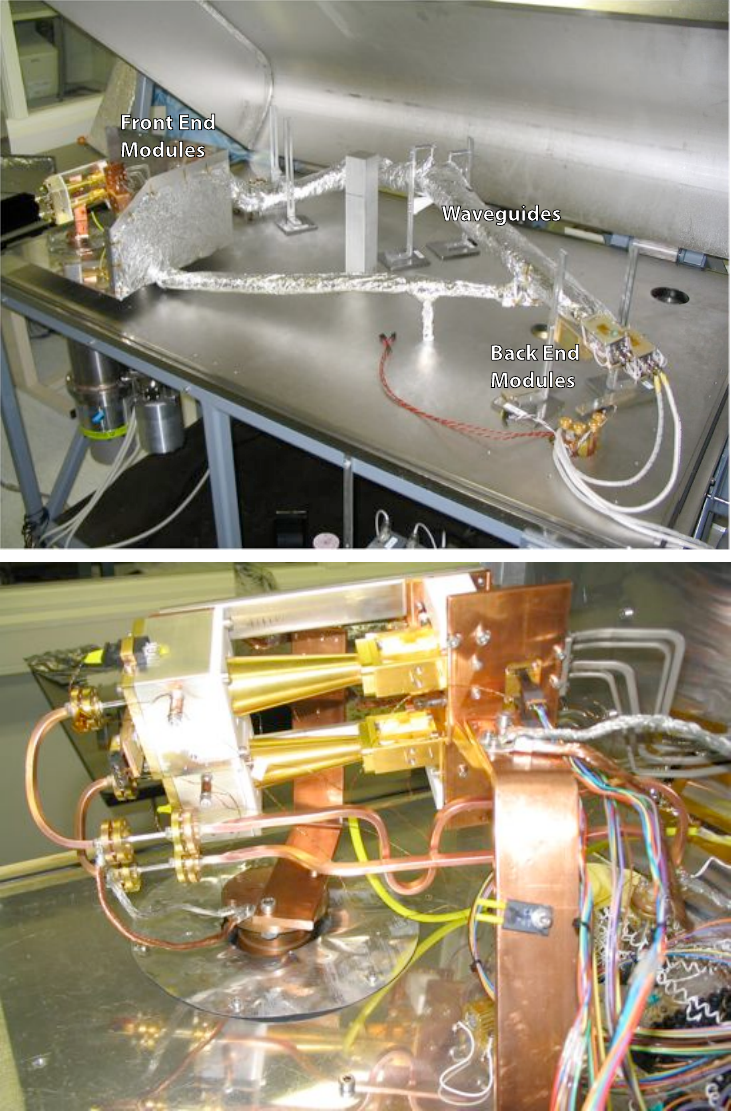} 
   \caption{Top: Two 70 GHz RCAs integrated in the Ylinen Electronics cryofacility. The two BEMs (on the right) are connected to the waveguides, here surrounded by aluminum mylar. On the left the shroud contains the two horns facing the ``Ylinen'' skyload at about 50K. Bottom: detail of the front end. The two FEMs and the pair of horns are facing the two skyload containers.}
   \label{fig:facility70}
\end{figure}

This cryofacility has the dimensions 1.6 $\times$ 1.0 $\times$ 0.3 $m^3$.
The facility has a layout similar to that at 30-44 GHz, although the 70 GHz facility was designed to house two radiometer chains simultaneously (Fig. \ref{fig:facility70}).\\
The smaller dimensions of the feedhorns and front end modules and the decision to use two small dedicated sky loads directly in front of the horns allowed the cold part of the two RCAs under test to be contained in a volume similar to that of the 30 and 44 GHz chamber.
Temperature interfaces such as FEMs, sky load and reference load were coupled together by means of copper slabs and then connected to the 4K and 20K coolers. 
The FEMs were controlled at their nominal temperature of 20 K; sky and reference loads were controlled in the range 10 -- 25 K with a stability better than 10 mK; the back end modules were insulated from the chamber envelope by means of a supporting structure without temperature control, which was considered unnecessary.

\subsection{The skyloads at 70 GHz}\label{sky70}
\begin{figure}
\centering
 \includegraphics[width=8.7cm]{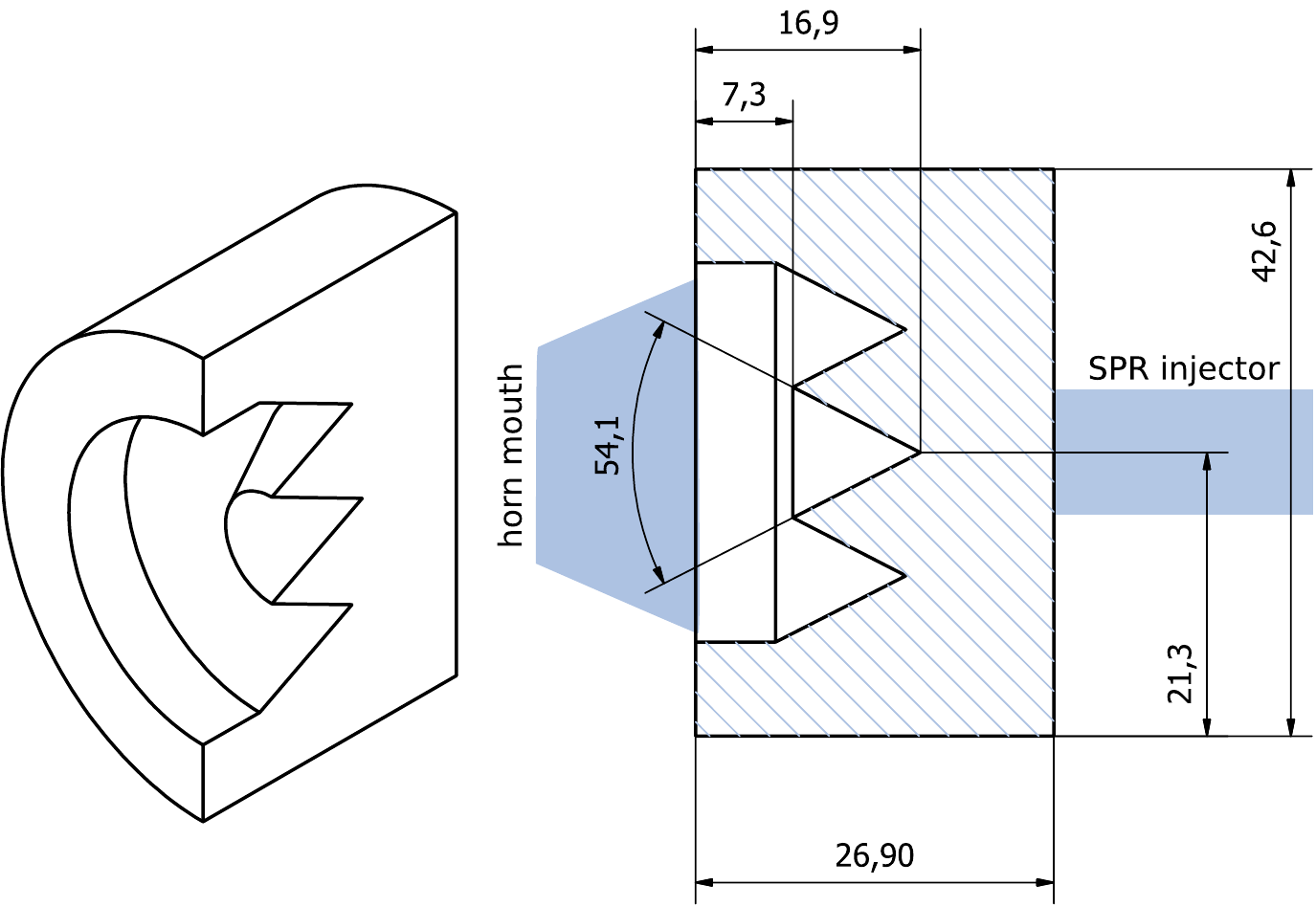} 
   \caption{Ylinen design of the RCA sky load calibrator. This design produced a load with $-20$dB of return loss over the whole bandwidth. The two light blue shaded regions represent the horn mouth at the left and the rectangular waveguide injector on the right. The absorber is enclosed in a metallic box except for the part facing the horn, which is closed with a Teflon\textregistered plate.}
 \label{fig:ylinen_sky_load}
\end{figure}

The design for the 70 GHz RCA skyload was made by Ylinen Electronics.  The basic design is shown in Fig. \ref{fig:ylinen_sky_load}.
The load configuration is a single folded conical structure in Eccosorb mounted in an aluminum housing. It is attached to a brass back plate. A single waveguide input is mounted through the back plate, providing the method of applying RF stimulus signals through the absorber for the {\tt RCA\_SPR} test (see Sect. \ref{spr}). Load performances were measured over the whole V--band showing a return loss better than $-20$dB.  
Two sensors were placed on the sky load, one at the controller stage, referred to as $T_{ctrl}$ and one inside the absorber, $T_{sky}$. Although the temperature along the skyload was expected 
to be uniform due to its small dimensions, this was not the case:
due to the cool down effects, the thermal junction between the temperature control and the load was not efficient as expected. A typical difference in temperature within $4-7$K was observed between the two thermometers. This cool-down effect was not predictable so that the sky load was considered as a relative instead of an absolute temperature reference.

\section{Methods and results}\label{results}

\subsection{Functional tests}
\begin{figure}
\centering
 \includegraphics[width=8.7cm]{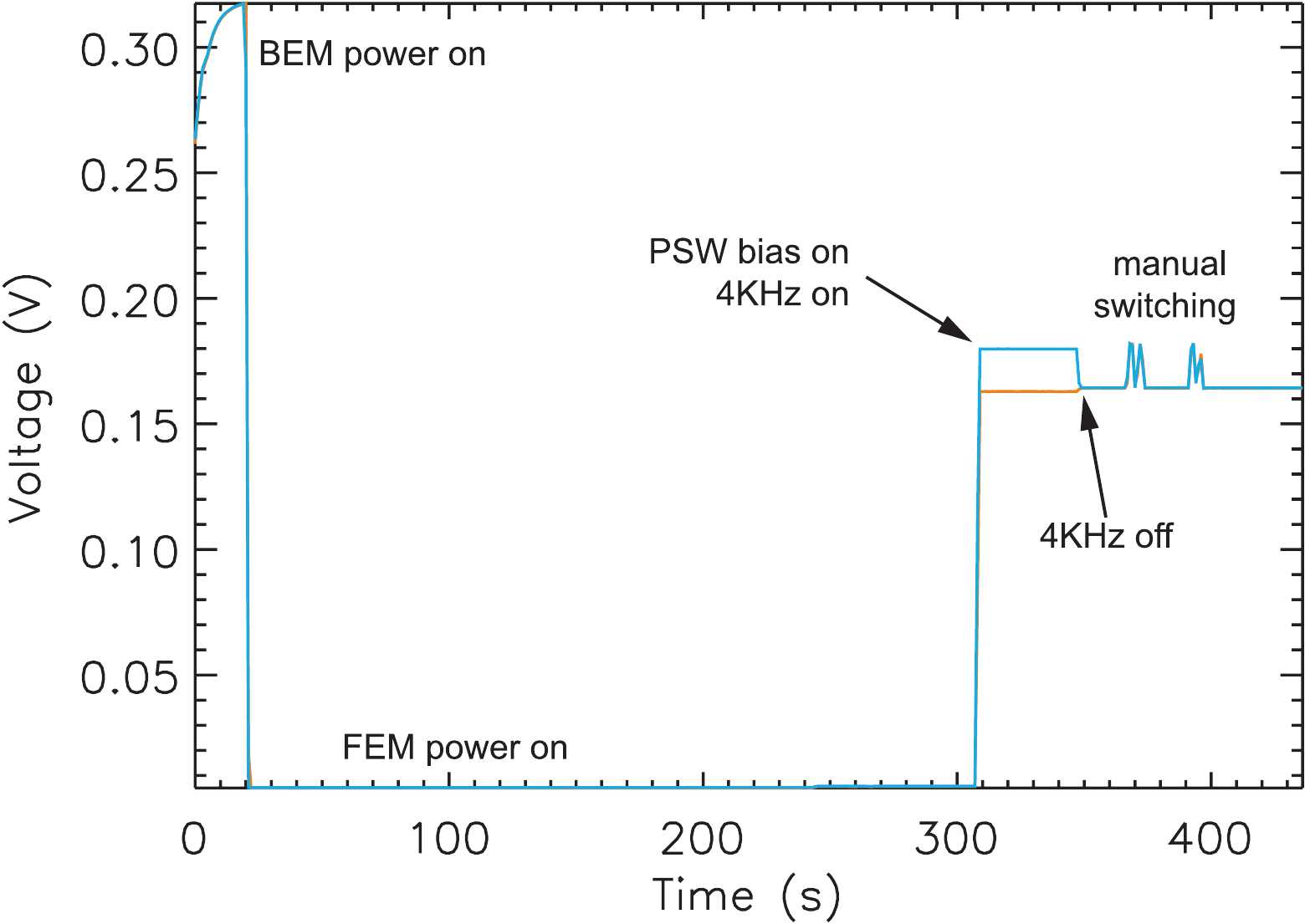} 
   \caption{Functional test performed at cryogenic temperature on {\tt RCA26}.  The two lines (orange and cyan) refer to the sky and load signals, when the 4KHz switching is activated. Outside the interval 310--350 seconds the two curves are indistinguishable.
}
   \label{fig:functional_tests}
\end{figure}

Functional tests were performed at ambient and at cryo temperature. All the RCAs were biased with nominal values and the power consumption was verified. In addition, each phase switch was operated in the nominal mode to check its functionality. 
As an example the functional test performed at cryogenic temperature on {\tt RCA26} is shown in Fig. \ref{fig:functional_tests}. 
The figure reports the output voltage of the detector {\tt S-10} when the functional test is run: the BEM is switched on, the FEM is biased at nominal conditions and the fast 4KHz switching is activated on the phase switches. It is evident that most of the change in signal is experienced when the BEM is on and the phase switches are biased correctly. 
These functional tests were also used as a reference for further tests up to the satellite-level verification campaign, besides checking the functionality of the RCA to proceed with the calibration.

\subsection{Tuning}
Before tuning the RCA, the DAE was set-up to read the voltage from each detector of the BEM, $V_{BEM}$, with appropriate resolution . The output signal from the DAE, $V_{DAE}$, is given by
 \begin{equation}
V_{DAE} = G^{DAE}\cdot \left ( V_{BEM} - O_{DAE}\right ).
\end{equation}

The DAE gain, $G^{DAE}$, was set to ensure that the noise induced by the DAE did not influence the noise of the radiometric signal from the BEM detectors. The voltage offset, $O_{DAE}$  was adjusted to guarantee that the output voltage signal lay well within the $[-2.5, +2.5]$ Volts range when the gain was set properly for the input temperature range.  $G_{DAE}$ and $O_{DAE}$ were set for each of the four detectors and employed during all noise property tests.

The aim of the RCA tuning procedure ({\tt RCA\_TUN}) was to choose the best bias conditions for each FEM low noise amplifiers' (LNA) gate voltage and phase switch current. Each of the four LNAs in a 30 GHz FEM consists of four amplification stages (five for the FEM at 44 GHz), each driven by the same drain voltage, $V_d$. 
The gate voltage $V_{g1}$ biases the first amplification stage, while $V_{g2}$  biases the successive three (or four) stages.  The phase switches are driven by two currents ($I_1$ and $I_2$), biasing each diode. The currents determine the amount of attenuation by each diode and thus are adjusted to obtain the final overall radiometer balance.

\begin{figure}
\centering
 \includegraphics[width=8.7cm]{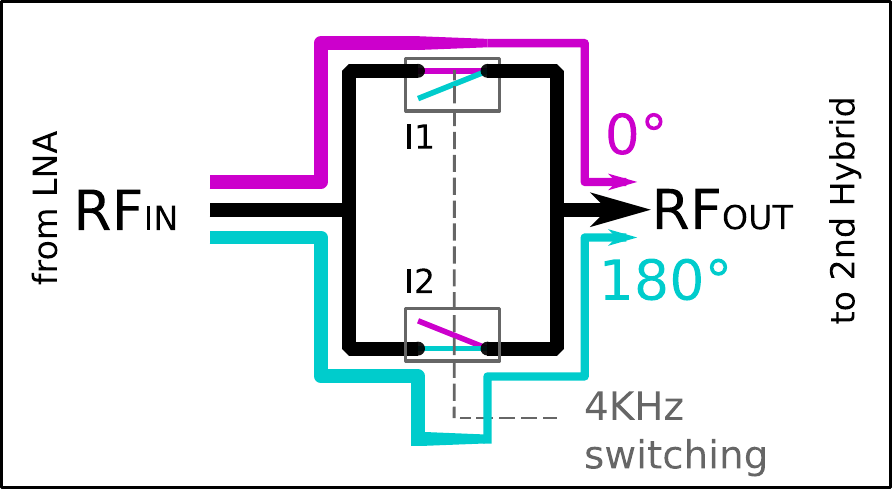} 
   \caption{Conceptual scheme of the phase switch integrated into the radiometers. Each phase switch is composed of two diodes commanded by the currents I1 and I2. They act as a on/off switch. Depending on the polarization state of the diodes the signal follows the magenta path or the cyan $\lambda/2$ shifted path. The two currents at  which the diodes are tuned determine the attenuation of each path, represented here by the different thickness. While the phase matching depends on the particular RF design,  the amplitude matching depends on the $(I_1,I_2)$ bias supply of the diodes which is the goal of the phase switch tuning.}
   \label{fig:psw-tuning}
\end{figure}

The phase switches between the LNA and the second hybrid use 
 the interconnection of two hybrid rings to improve the bandwidth and the matching with two Shunt PIN diodes. Depending on the polarization of the diodes the signal travels into a circuit, which can be $\lambda$ / 2 longer, so that it is shifted by $180^\circ$. Details can be found in \citet{hoyland2004} and in \citet{cuttaia2009}. In Fig. \ref{fig:psw-tuning} we report the conceptual schematic diagram of the phase switch.
 
At 30 and 44 GHz the phase switches were tuned with one radiometer leg switched off. In these conditions the signal entering each phase switch diode is the same, and the output signal at the DAE can be used directly to precisely balance the two states of the phase switch. Any differences in the sky and ref signal are caused only by the phase switches and not by other non-idealities of the radiometers, nor by different input target temperatures.  The two currents were chosen to minimize the quadratic differences, $\delta^{PSW}$,  between odd and even samples of the signal (corresponding in Fig. \ref{fig:psw-tuning} to the magenta and cyan path respectively). If for example the phase switch was tuned on the same leg as the amplifier M1, the following expression was minimized: 

\begin{equation}
\delta^{PSW}_{M1} = \sqrt{\left ( S_{00}^{o}-S_{00}^{e}\right ) ^2+ \left ( S_{01}^{o}-S_{01}^{e}\right ) ^2},
\end{equation}
where $S_{00}$, $S_{01}$ are the two DAE outputs related to the 'M' half FEM. In this case $o$ and $e$ refer to odd and even signal samples. The same differences for the other phase switches, $\delta^{PSW}_{M2}$, $\delta^{PSW}_{S1}$, and $\delta^{PSW}_{S2}$ were calculated in the same way. The $I_1$ and $I_2$ were varied around the best value obtained during the FEM stand-alone tests \citep{davis2009}. The phase switches of the 70 GHz RCAs were not tuned.  To reduce the transient, the phase switches were always biased at the maximum allowable current.

The front-end LNAs were tuned for noise temperature performance, $T_n$, and isolation, $I$. For each channel $T_n$ and $I$ were measured as a function of the gate voltages $V_{g1}$ and $V_{g2}$.

Firstly the minimum noise conditions were found by varying $V_{g1}$ while keeping $V_{g2}$ and $V_{d}$ fixed.  The noise temperature was measured with the Y-factor method (see Appendix \ref{a:Y-factor} for the details of this method). Because only relative estimates of Tn are relevant for tuning purposes, we did not correct for the effect of non-linearity in the 30 and 44 GHz RCAs. 

Once the optimum  $V_{g1}$ was determined, the optimum $V_{g2}$ was found by maximizing the isolation, $I$, 
\begin{equation}
I = { \Delta V_{sky} - G_0\cdot\Delta T_{sky} \over \left ( \Delta V_{ref} - G_0\cdot\Delta T_{sky}\right ) +  \Delta V_{sky} },
\label{eq:isolation}
\end{equation}

measured by varying  $T_{ref}$ with $T_{sky}$ kept fixed. The term $-G_0\cdot\Delta T_{sky}$ is a correction factor for any unwanted variation of $T_{sky}$ originating chiefly from thermal non-ideality of the cryofacility. 

At 70 GHz the best working conditions were found by measuring the
noise temperature and the isolation as a function of the gate voltages
$V_{g1}$ and $V_{g2}$, but with a slightly different approach mainly for schedule reasons to that of the lower frequency chains. The procedure required two different temperatures for the
reference load (about 10 K in the low state and about 20 K in the high
state) and $V_{g1}$, $V_{g2}$, and $V_{d}$ were varied independently on the half FEM. Then the procedure was repeated with both FEM legs switched on. The noise temperature was measured with the Y-factor method from the signals coming from the two temperature states, and the isolation was calculated with Eq. (\ref{eq:isolation}).

Although the bias parameters found during the RCA tuning are the optimal ones, we found that different electrical and cryogenic conditions induce uncertainties in the bias values. This is due to the different grounding and the impact of the thermal gradient along the bias cables. To overcome this problem, tuning verification campaigns are planned at LFI integrated level and inflight. In both cases the RCA bias values have been assumed as reference values.

\subsection{Basic properties}\label{basic_properties}
The basic properties of the radiometers, namely noise temperature, isolation, gain,  and linearity were obtained in a single test.

\begin{figure}
\centering
\includegraphics[width=8cm]{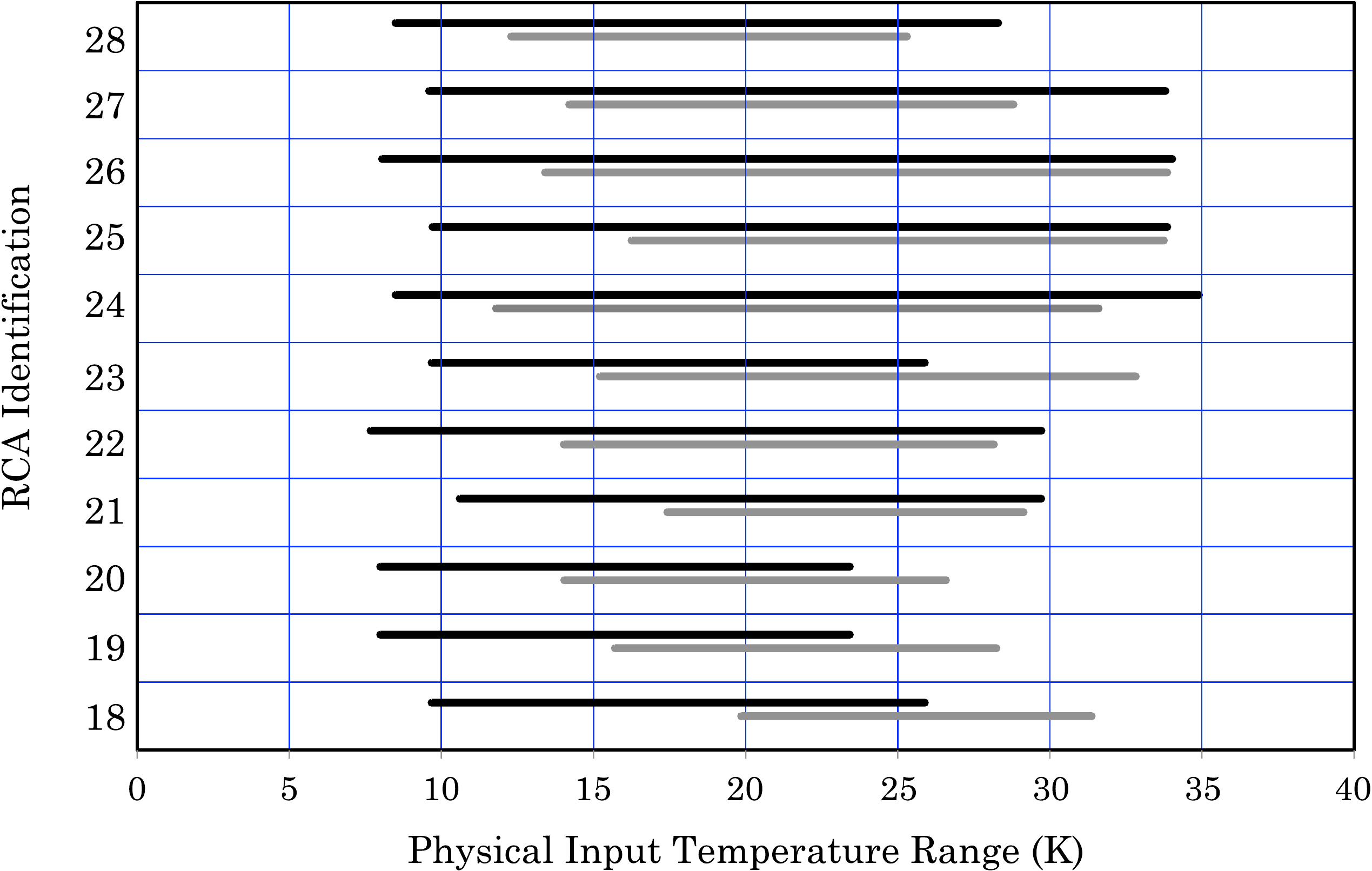} 
   \caption{Physical temperature ranges for the {\tt RCA\_LIS} tests. Black lines refer to the reference load temperature steps. Gray lines refer to the skyload temperature steps.}
   \label{fig:t_range}
\end{figure}

The {\tt RCA\_LIS} test was performed varying $T_{sky}$ (and subsequently $T_{ref}$) in steps while keeping the $T_{ref}$ (and subsequently $T_{sky}$) constant. In Fig. \ref{fig:t_range} we give the temperature range spanned during the tests. Due to the different thermal behavior of each RCA the range was not the same for all the chains. The brightness temperature was calculated from temperature sensors  located in the external calibrators (both sky and reference) and the output voltages of the four detectors were recorded. The main uncertainty was in the determination of the actual brightness temperature seen by the radiometer. At 30 and 44 GHz the brightness temperature of the skyload was derived from the thermometer located inside the pyramids, from where the main thermal noise emission originated. For 70 GHz both the backplate thermometer and the absorber thermometer were used to derive the brightness temperature. However,  the backplate and the absorber temperatures were found to introduce a significant systematic error in the reconstructed physical temperature of the load, as explained in Sect. \ref{sky70}. It was decided to calibrate the 70 GHz RCAs using the reference load steps instead.

We denote here for simplicity the value of either $V^{sky}_{out}$ or $V^{ref}_{out}$ with $V_{out}$  in Eqs. (\ref{eq:sky1}) and (\ref{eq:ref1}),  alternatively $T_{sky}$ or $T_{ref}$ with $T_{in}$, and the corresponding $\widetilde{T}^{sky}_N$ or $\widetilde{T}^{ref}_N$ with $T_{sys}$. With $G$ we denote the corresponding total gain.  

For a perfectly linear radiometer the output signal can be written as
\begin{equation}
V_{out} = G\cdot (T_{in} + T_{sys}),
\end{equation}
and the gain and system temperature can be calculated by measuring the output voltage for only two different values of the input temperature (Y--factor method). This was indeed the case for the 70 GHz RCAs. For the 30 and 44 GHz RCAs the determination of the basic properties was complicated by a significant non-linear component in the response of the 30 and 44 GHz RCAs. This was discovered during  the previous qualification campaign and has been well characterized during these flight model RCA tests. The non-linearity effects in LFI are discussed in \citet{mennella2009b} together with its impact on the science performances.  For a radiometer with compression effects the radiometer gain, $G$, is a function of total input temperature, $T = T_{in} + T_{sys}$ and is given by
\begin{equation}
\label{eq:nl}
V_{out} = G(T) \cdot (T_{in} + T_{sys}) .
\end{equation}
Particular care was required in the determination of the noise temperature of the 30 and 44 GHz RCAs. To overcome the problem, the application of four different types of fit were performed: 

\begin{enumerate}
\item  {\it linear fit.} This fit was always calculated as a reference, even for non-linear behavior of the radiometer,  so that
\begin{equation}
V_{out} = G_{lin}\cdot T.
\end{equation}
The linear gain, $G_{lin}$, and the noise temperature were derived. The fit was applied to all available data, not only to the two temperature steps as in the Y-factor method, to reduce the uncertainties for the linear 70 GHz RCAs and to evaluate the non-linearity of the 30 and 44 GHz chains.

\item {\it parabolic fit.} This was applied to understand the effect of the non-linearity, for the evaluation of which the quadratic fit is the simplest way. The output of the fit were the three coefficients from the equation
\begin{equation}
V_{out} = a_0 + a_1 T + a_2 T^2.
\end{equation}
In this case the noise temperature was employed as the solution of the equation $a_2 T_{sys}^2 + a_1 T_{sys} + a_0= 0$. 
\item {\it inverse parabolic fit.} This was used because the noise temperature was directly derived from 
\begin{equation}
T = a_0 + a_1 V_{out} + a_2 V_{out}^2,
\end{equation}
where $T_{sys} = a_0$. 

\item {\it gain model fit.} A new gain model was developed based on the results of \citet{daywitt1989}, modified for the LFI (see Appendix \ref{app:gain-model}). The total power output voltage was written  as
\begin{equation}
V_{out} = \left [ {G_0 \over 1 + b \cdot G_0 \cdot \left (T_{in} + T_{sys}\right ) } \right ] \cdot \left (T_{in} + T_{sys}\right ),
\end{equation}
where $G_0$ is the total gain in the case of a linear radiometer, $b$ is the linear coefficient ($b=0$ in the linear case, $b=\infty$ for complete saturation, i.e. $G(T) =0$).  
\end{enumerate}

The values obtained for gain, linearity and noise temperature are reported in Table \ref{tab:basic_performances}. The isolation values were calculated with Eq. (\ref{eq:isolation}) based on all possible combinations of temperature variation on the reference load. The results are given in Table \ref{tab:isolation}, where  the values obtained during the tuning of the $V_{g2}$ are also reported for the 30 and 44 GHz RCAs. 
While at 30 GHz the differences are due mainly to the reference load thermal model applied in this case, for the 44 GHz RCAs the differences are dominated by the gain used to compensate for thermal coupling in Eq.  (\ref{eq:isolation}). 

\begin{table}
\caption{Receiver basic properties.  Isolation in dB. The values found during the $V_{g2}$ tuning are also reported in brackets for comparison.  
}             
\label{tab:isolation}      
\centering                          
\begin{tabular}{l r r r r}        
\hline\hline                 
 & \multicolumn{4}{c}{\sc{Isol ($dB$)}}\\
 & \multicolumn{4}{c}{\sc{Linear Model}} \\
 & M-00 & M-01 & S-10 & S-11 \\    
\hline                        
{\tt RCA18}  & -13.5	& -13.0	& -11.1	& -10.9 \\
{\tt RCA19}  & -15.5	& -15.9	& -15.3	& -13.9 \\
{\tt RCA20}  & -15.8	& -15.9	& -12.7	& -14.2 \\
{\tt RCA21}  & -13.0	& -12.4	& -10.1	& -10.4 \\
{\tt RCA22}  & -12.8	& -11.1	& -12.4	& -11.3 \\
{\tt RCA23}  & -12.6	& -11.8	& -13.3	& -14.3 \\
\hline
{\tt RCA24}  & -11.7  & -12.3 & -10.4 & -10.5 \\
             & (-13.3)	& (-13.6)& (11.6)& (-11.9)\\
{\tt RCA25}  & -10.8 	& -10.7 	& -12.0  & -11.5  \\
& (-14.6)	&(-14.4)	& (-15.5) & (-14.5) \\
{\tt RCA26}  &  -10.8  & -11.9 & -13.7 & -13.7 \\
&  (-9.7) &  (-10.4)& (-13.9)& (-14.0)\\
\hline
{\tt RCA27}  & -13.0 & -12.8 	& -14.7 	& -14.6 \\
&  (-11.2)	&  (-11.0)	&(-11.7)	& (-11.8)\\
{\tt RCA28}  & -10.9  & -10.3  & -10.3 & -10.5   \\
& (-11.6) &(-11.2) & (-12.0) & (-12.2)  \\
\hline                                   
\end{tabular}
\end{table}
%

\begin{table}
\caption{Receiver basic properties: gain, noise temperature, and linearity. For the $70$GHz RCAs, the gain, $G_0$, and the noise temperature, $T_n$, were derived from the linear fit, and the linearity coefficient is not reported. For $30$ and $44$ GHz, $G_0$, 
$T_n$, and the linearity coefficient, $b$, were derived from the gain model-fit.}            
\label{tab:basic_performances}      
\centering                          
\begin{tabular}{l l r r r r}        
\hline\hline                 
 & & \multicolumn{4}{c}{\sc{Gain (V/K), $T_{n}$ (K), and Lin}}\\
 & & M-00 & M-01 & S-10 & S-11 \\    
\hline                        
{\tt RCA18}  & $G_0$ & 0.0173 &	0.0195 &	0.0147	& 0.0143 \\
             & $T_{n}$  & 36.0 & 36.1& 33.9 & 35.1 \\
             \hline
{\tt RCA19}  & $G_0$ & 0.0161 &	0.0174 &	0.0176 &	0.0196 \\
             & $T_{n}$ & 33.1 & 31.5 & 32.2 & 33.6 \\
             \hline
{\tt RCA20}  & $G_0$ & 0.0186 &	0.0164 &	0.0161 &	0.0165\\
             & $T_{n}$ & 35.2 & 34.2 &	36.9 & 35.0 \\
\hline
{\tt RCA21}  & $G_0$ & 0.0161 &	0.0154 &	0.0119 &	0.0114\\
             & $T_{n}$ & 27.3 & 28.4 & 34.4 & 36.4 \\
\hline
{\tt RCA22}  & $G_0$ & 0.0197 &	0.0174 &	0.0165 &	0.0163 \\  
             & $T_{n}$ & 30.9 & 30.3 & 30.3 & 31.8 \\
{\tt RCA23}  & $G_0$ & 0.0149	& 0.0171 &	0.0271 &	0.0185\\
             & $T_{n}$ & 35.9 & 34.1 & 33.9 & 31.1 \\
             \hline
{\tt RCA24}  & $G_0$   & 0.0048 &	0.0044 &	0.0062 &	0.0062 \\
             & $T_{n}$ & 15.5	& 15.3 &	15.8 &	15.8 \\
             & $b$     &1.79	& 1.49 &	1.44 &	1.45 \\
             \hline
{\tt RCA25}  & $G_0$   & 0.0086	& 0.0085 &	0.0079 &	0.0071 \\
						 & $T_{n}$ & 17.5	& 17.9 &	18.6 &	18.4\\

             & $b$     & 1.22	& 1.17 &	0.80 &	1.01\\
             \hline
{\tt RCA26}  & $G_0$   & 0.0052	& 0.0067 &	0.0075 &	0.0082\\
						 & $T_{n}$ & 18.4	& 17.4 &	16.8 &	16.5 \\
             & $b$     & 1.09	& 1.42 &	0.94 &	1.22\\
             \hline
{\tt RCA27}  & $G_0$   & 0.0723	& 0.0774 &	0.0664 &	0.0562\\
						 & $T_{n}$ & 12.1	& 11.9 &	13.0 &	12.5\\
             & $b$     & 0.12	& 0.12 &	0.13 &	0.14\\
             \hline
{\tt RCA28}  & $G_0$   & 0.0621 &	0.0839 &	0.0607 &	0.0518\\
						 & $T_{n}$ & 10.6	& 10.3 &	9.9 &	9.8\\
             & $b$     & 0.19	& 0.16 &	0.19 &	0.20\\
\hline                                   
\end{tabular}
\end{table}

\subsection{Noise properties}\label{noise_properties}
\begin{figure}
\centering
\includegraphics[width=8cm]{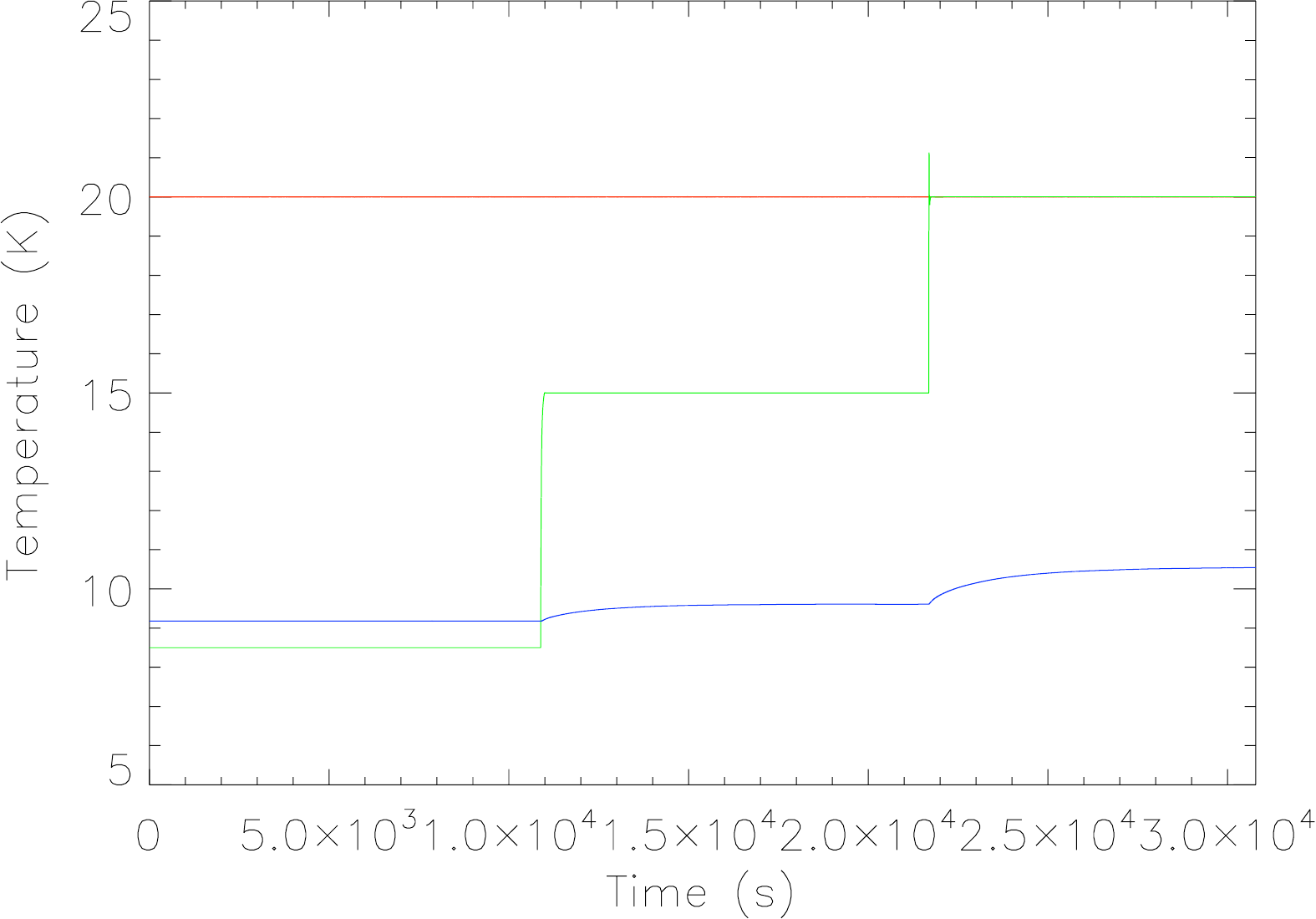} 
   \caption{Typical temperature behavior of sky load (blue), reference load (green) and FEM body (red) during the {\tt RCA\_STn} test. The temperature stability is better than $1$mK.}  
\label{fig:st3_temp}
\end{figure}

\begin{figure*}
\centering
\begin{tabular}{l c r}
\includegraphics[width=5.6cm]{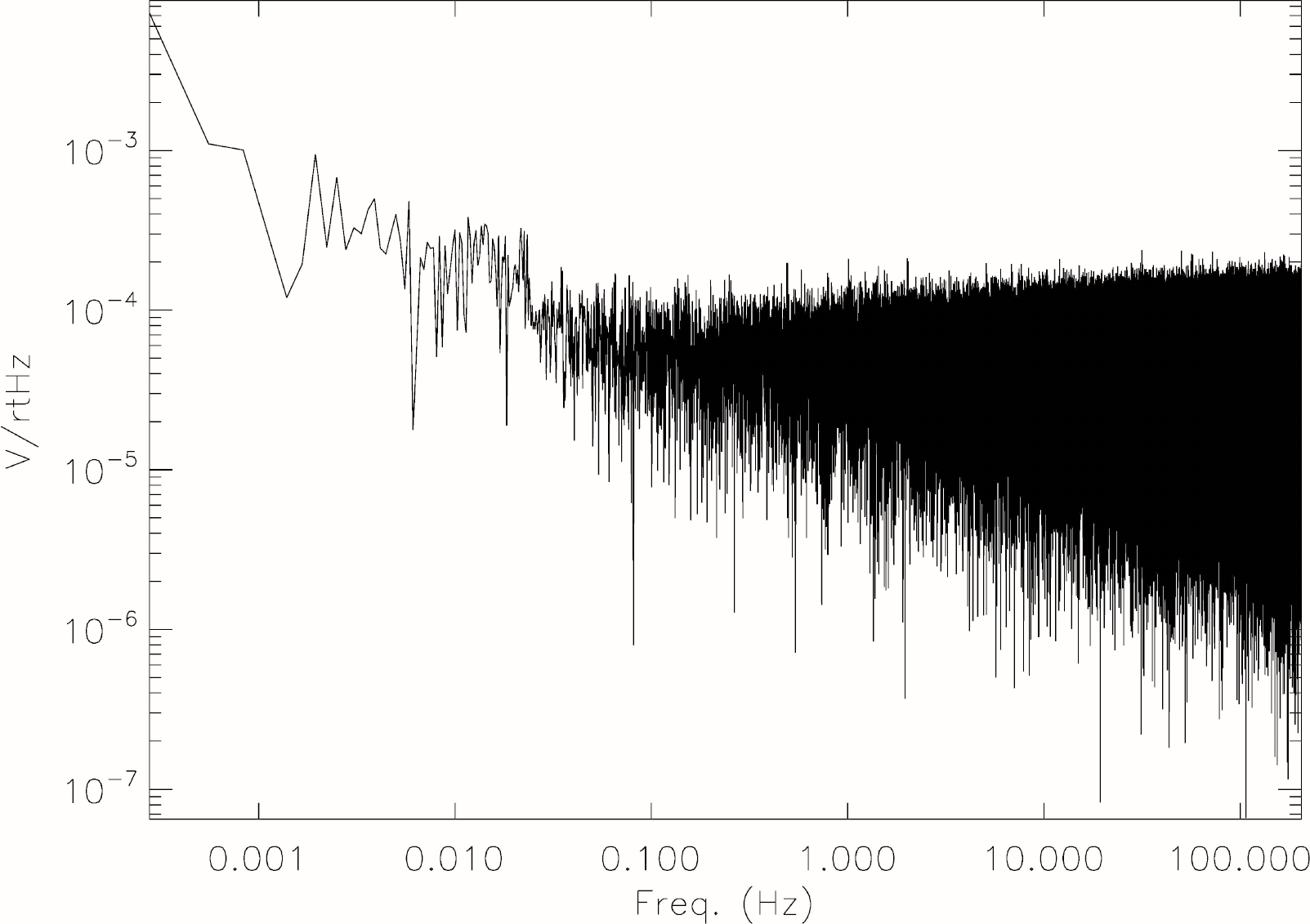} & \includegraphics[width=5.6cm]{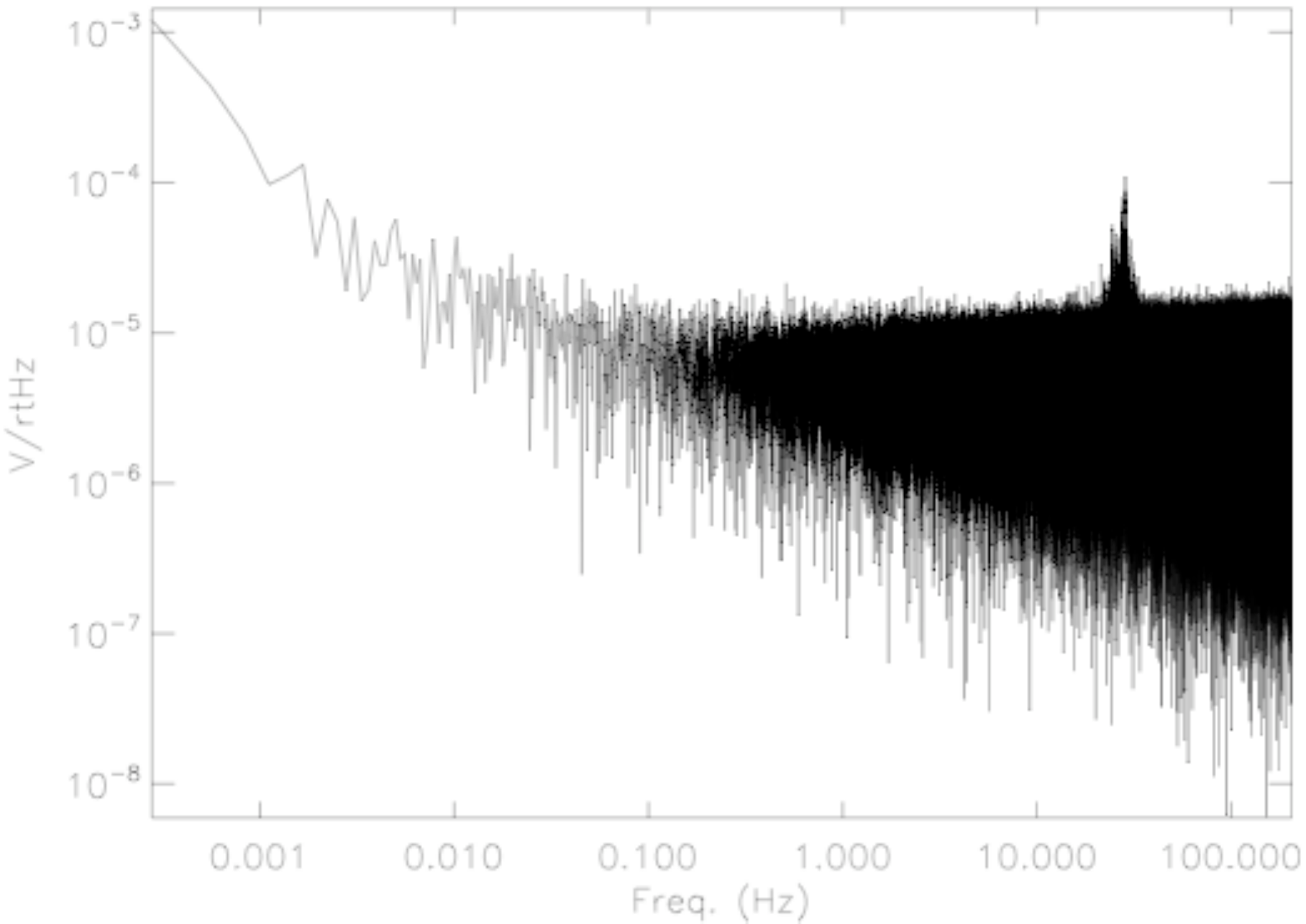} & 
\includegraphics[width=5.6cm]{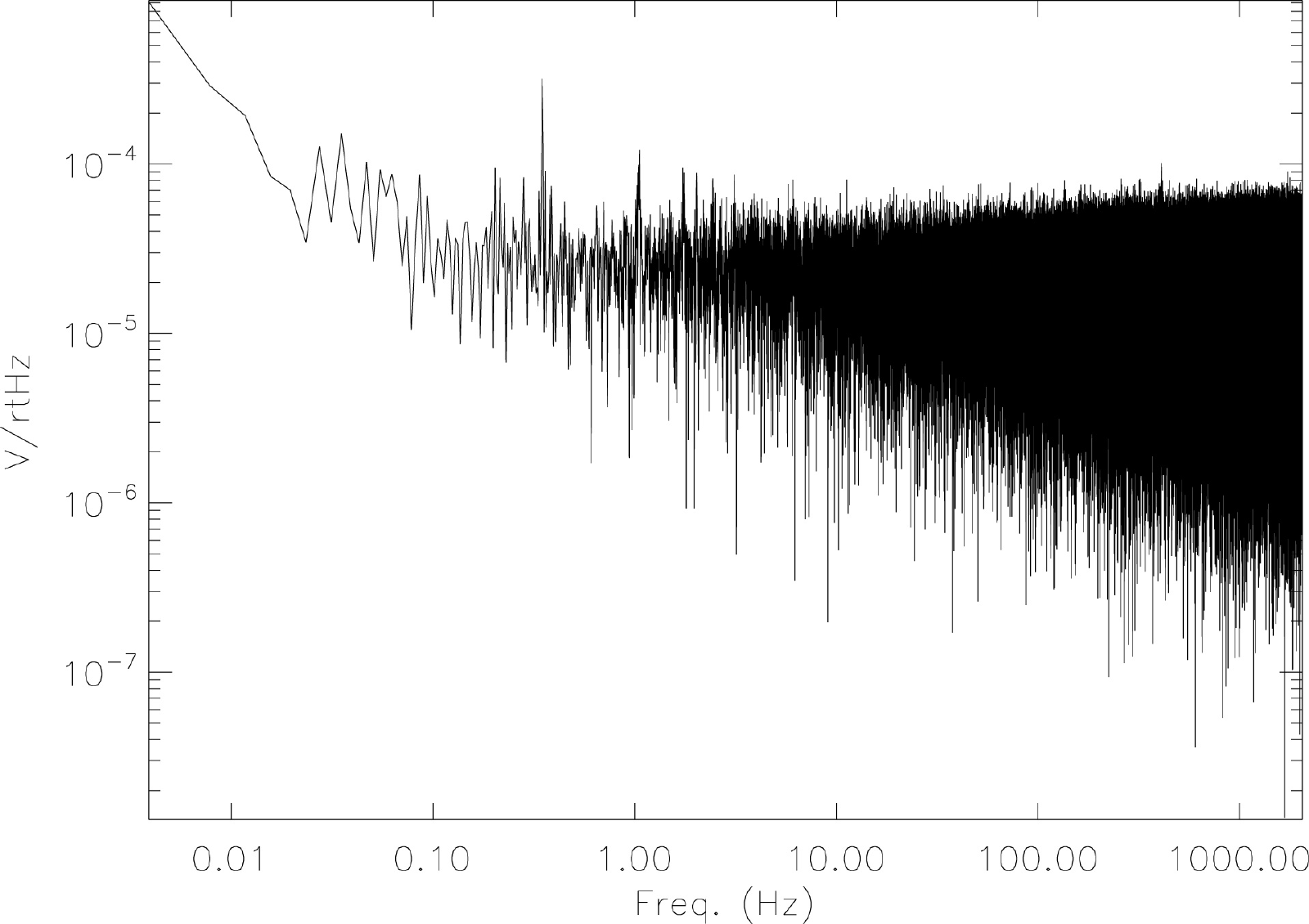} \\
\end{tabular}
   \caption{Log-log plot of the amplitude spectral density  of the differential detector output noise. The left plot refers to the {\tt RCA27M-00} detector with the sky and the reference loads both at 20K; the plot in the center refers to the {\tt RCA26M-00}  detector with sky and reference loads at 8K and 13K respectively; the plot on the right refers to the {\tt RCA23S-11} with the sky and reference loads at 15K and 9K respectively.}
\label{fig:fft}
\end{figure*}

Radiometer noise properties were derived from the {\tt RCA\_STn} test.  This test consisted of acquiring data under stable thermal conditions for at least three hours. Then the temperature of the loads were changed to measure the noise properties at different sky and reference target  temperatures. As expected, the best 1/f conditions were found  when the difference between the sky and reference load temperatures was minimal. This occurred at the first and last step as seen in Fig. \ref{fig:st3_temp}, which represents a typical temperature profile of the test.
The amplitude spectral density was calculated for each output diode. The 1/f component (knee frequency, $f_k$, and slope, $\alpha$), the white noise plateau and the gain modulation factor, $r$, were derived. 
At 70 GHz the 1/f spectrum was clearly dominated by thermal instabilities of the BEM which was not controlled in temperature. There the measured knee frequency values were over-estimated, while at 30 and 44 GHz the cryofacility was sufficiently stable to characterize the 1/f performances of the radiometers.  From the white noise and DC level the effective bandwidth was  calculated as
\begin{equation}
\beta = K^2 \cdot {V_{out}^2\over \Delta V_{out}^2 \cdot \tau},
\end{equation}
where $\Delta V_{out}$ is the white noise level, $V_{out}$ is the DC level, $\tau$ the integration time, and $K=1$ for a single detector total power in the unswitched condition, $K=\sqrt2$ for a single detector total power in the switched condition, $K=2$ for a single detector differenced data, $K=\sqrt2$ for double-diode differenced data. This formula does not include the non-linearity effects that are discussed in detail by \citet{mennella2009b}. The overall noise performances of all eleven RCAs are reported in Table \ref{tab:noise_performances}, while Fig. \ref{fig:fft} shows the typical amplitude spectral density  of the noise for each frequency channel.

\begin{table}
\caption{Receiver noise properties: 1/f knee frequency, slope, and r factor. Data were taken setting the temperature of the loads at the lowest possible values. The 70 GHz 1/f knee frequencies are dominated by the thermal instabilities of the cryochamber.} 
\label{tab:noise_performances}      
\centering                          
\begin{tabular}{l l r r r r}        
\hline\hline                 
 & & \multicolumn{4}{c}{\sc{1/f spectrum, r factor, and $\beta$}}\\
 & & M-00 & M-01 & S-10 & S-11 \\    
\hline                        
{\tt RCA18}  
& $f_k$ &{\bf 92} &{\bf 92} & {\bf 140} & {\bf 190}\\
& $\alpha$& -1.62 & -1.71 & -1.49 & -1.71 \\
& $r$    & 1.17& 1.17& 1.17& 1.16 \\
& $\beta$  & 9.57   & 10.30  &  8.53 & 10.78\\
\hline
{\tt RCA19}  
& $f_k$  & 130 &	96&144	&143 \\
& $\alpha$& {\bf -1.75 }& {\bf -1.93 }& {\bf -1.80} & {\bf -1.85}\\
& $r$    & {\bf 1.12}& {\bf 1.12}& {\bf 1.14}& {\bf 1.13 } \\
& $\beta$  &  8.90  &  9.10 & 9.00  & 11.06\\
\hline
{\tt RCA20}  
& $f_k$  & 55.7&	63.7&	119.3& 98.5\\
& $\alpha$&-1.63 &-1.48 &-1.36 & -1.32\\
& $r$    & 1.03& 1.02& 1.03& 1.04 \\
& $\beta$  & 10.74   & 9.46  & 10.86  & 10.48\\
\hline
{\tt RCA21}  
& $f_k$  & 109&	85&	109& 97\\
& $\alpha$& -1.59& -1.90&-1.61 &-1.17 \\
& $r$    & 1.21&1.21 &1.18 &1.18  \\
& $\beta$  &10.02    &10.08   & 9.79  &8.79 \\
\hline
{\tt RCA22}  
& $f_k$  &116 &108	&80	& 113\\
& $\alpha$& -1.59& -1.90& -1.61& -1.65\\
& $r$    & 1.21& 1.21& 1.18& 1.18 \\
& $\beta$  &   10.02 &10.8   & 9.79  &8.79 \\
\hline
{\tt RCA23}  
& $f_k$  & 97&89	&101	&109 \\
& $\alpha$& -1.41& -1.92& -1.90& -1.83\\
& $r$    & 1.07&1.07 &1.08 &1.08  \\
& $\beta$  &  11.31  & 13.02  & 11.05  & 11.89\\
\hline
{\tt RCA24}$^\mathrm{a}$
& $f_k$  &13.9 &10.2&9.7& 13.3\\
& $\alpha$&-1.13& -1.26&-1.15& -1.18\\
& $r$    &0.991&0.974&0.971& 0.962\\
& $\beta$  &   6.13 &  4.12 &  5.21 & 6.59\\
\hline
{\tt RCA25}$^\mathrm{b}$
& $f_k$  &18.0&	21.9&	9.6& 4.7\\
& $\alpha$& -1.12&-1.28&1.04&-0.89\\
& $r$    &1.0289& 1.059& -0.859&   1.041\\
& $\beta$  & 6.87   & 6.88  & 4.96  & 6.82\\
\hline
{\tt RCA26}$^\mathrm{c}$
& $f_k$  &21.0&23.2&15.4&23.1\\
& $\alpha$&-1.02&-0.72&-0.67& -0.81\\
& $r$    &1.085&1.061&1.131& 1.119\\
& $\beta$  &  5.67  & 5.52  &   5.01& 7.40 \\
\hline
{\tt RCA27}$^\mathrm{d}$
& $f_k$  & 6.7&10.1&	24.9&31.1\\
& $\alpha$&-1.02&-1.19&-1.39& -0.90\\
& $r$    &1.012&1.004&1.093& 1.075\\
& $\beta$  & 7.77   & 7.70  & 8.73  & 7.18\\
\hline
{\tt RCA28}$^\mathrm{e}$
& $f_k$  &19.9&19.4&40.7&41.1\\
& $\alpha$& -1.39& -1.20& -1.57&-1.60\\
& $r$    &1.058 &1.050&0.955& 0.939\\
& $\beta$  &  7.91  & 7.94  &8.78   & 8.23\\
\hline       
                            
\end{tabular}
\begin{list}{a}{}
\item[$^{\mathrm{a}}$] $T_{ref}=8.5$K; $T_{sky}=8.5$K
\item[$^{\mathrm{b}}$] $T_{ref}=8.0$K; $T_{sky}=10.5$K
\item[$^{\mathrm{c}}$] $T_{ref}=8.0$K; $T_{sky}=13.0$K
\item[$^{\mathrm{d}}$] $T_{ref}=9.5$K; $T_{sky}=12.8$K
\item[$^{\mathrm{e}}$] $T_{ref}=8.6$K; $T_{sky}=8.5$K
\end{list}
\end{table}

\begin{table}
\caption{Receiver noise properties.  Long duration test antenna temperature pairs for the {\tt RCA28}. }          
\label{tab:noise_example}      
\centering                          
\begin{tabular}{ c c c c c}        
\hline\hline                 

 $T_{sky}$[K] &  \multicolumn{2}{c}{$T_{ref}$[K]} &  \multicolumn{2}{c}{$\left ( T_{ref} -T_{sky}\right )$} \\
                   & M  & S  & M & S \\    
                   \hline
 8.48 	&  10.21 & 8.85  	& 1.73  & 0.37  \\ 
 9.07 		&  15.16 & 14.66 	&  6.10 &  5.59\\
 9.74 		&  19.45 & 19.36 	&  9.70  &  9.62\\
 19.63 	&  19.45 & 19.36 	&  -0.18 &  -0.27\\
\hline                        
\end{tabular}
\end{table}

\begin{figure}
\centering
\includegraphics[width=9cm]{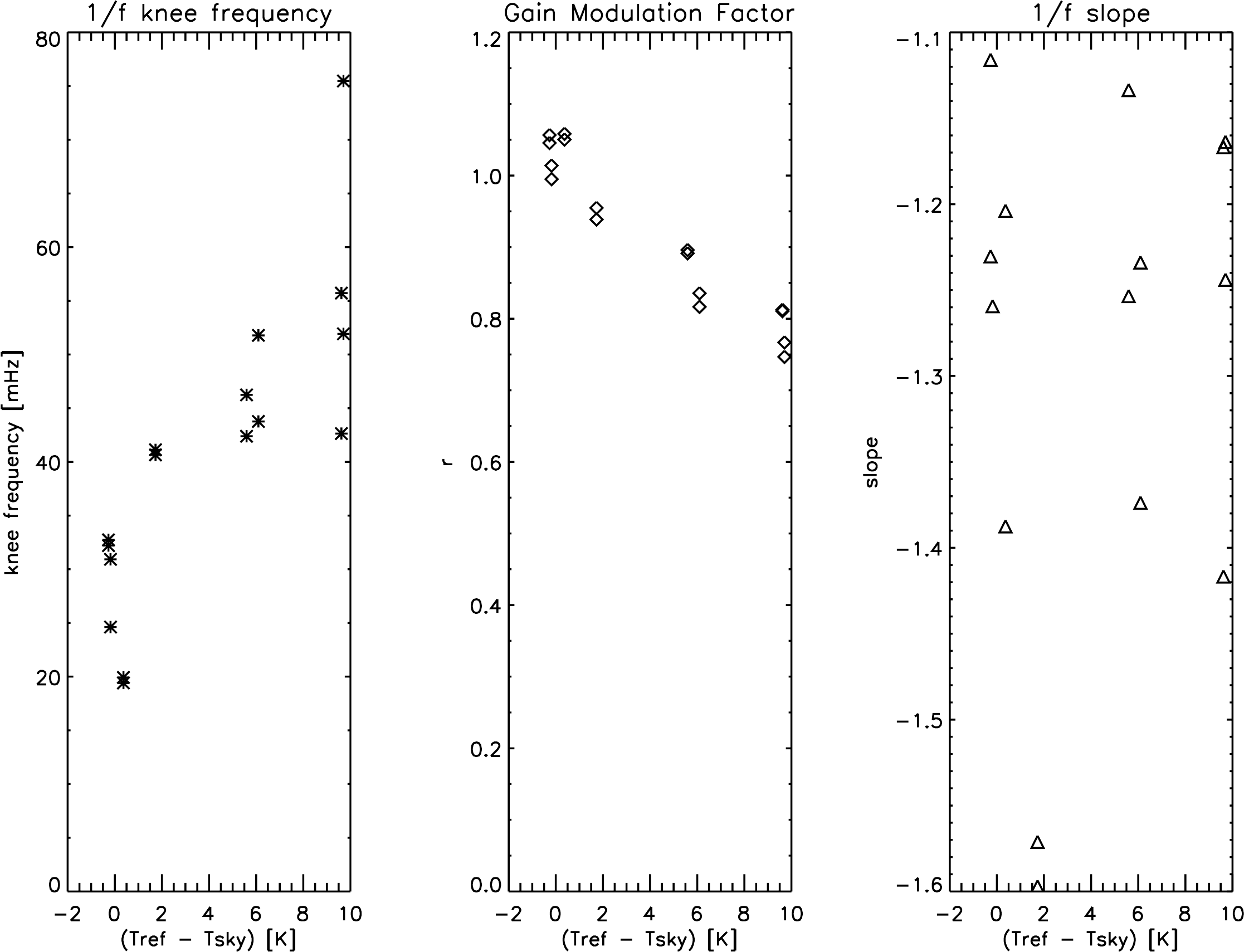} 
  \caption{1/f knee frequency (asterisk on the left), gain modulation factor (diamonds in the center), and the slope of the 1/f spectrum (triangles on the right) as a function of $\left (T_{ref}-T_{sky}\right )$. Sixteen points (four pairs for each detector) were reported. The spread of knee frequency values is due to the intrinsic difficulty of fitting the lower part of the power spectral density.  
}
\label{fig:rca28_noise}
\end{figure}
Apart from 1/f noise and white noise, spikes were observed in all RCAs: at $70$GHz they were caused by the electrical interaction between the two DAE units, which were slightly unsynchronised; at $30$ and $44$ GHz they were due to the housekeeping acquisition system. Because it was clear that the spikes were always due to the test setup and not to the radiometers themselves, the spikes were not considered critical at this stage, even if they showed up in frequency and amplitude.

As an example of the dependence of the noise performance on the temperature the antenna temperature pairs used during the  the tests of the {\tt RCA28} are reported in Table \ref{tab:noise_example}. These temperatures were calculated with the coefficients reported in Tables \ref{tab:Rfit} and \ref{tab:Tfit} of Sect. \ref{facility_3044} with the physical temperatures converted into antenna temperatures. The differences between the $T_{ref}$ and the $T_{sky}$ were calculated for each arm of the radiometer. The resulting 1/f knee frequency, the slope of the 1/f spectra, and the gain modulation factor, $r$, are reported in Fig. \ref{fig:rca28_noise} as a function of the temperature differences. 
It is evident from these plots that the knee frequency is increasing with the temperature difference, as expected. 
Moreover, the gain modulation factor is approaching unity as the input temperature difference becomes zero, which agrees with Eq. (\ref{eq:r}). 
The slope, $\alpha$, does not show any correlation with the temperature differences, because it depends on the amplifiers
rather than on the $\left ( T_{ref}-T_{sky}\right )$. This behavior was also found in the other RCAs. 

\subsection{Bandpass}\label{spr}
A dedicated end-to-end spectral response test, {\tt RCA\_SPR}, was designed and carried out to measure radiometer RF bandshape in operational conditions, i.e., on the integrated RCA with the front-end at the cryogenic temperature.  An external RF source was used to inject a monochromatic signal sweeping through the band into the sky horn. Then the DC output of the radiometer was recorded as a function of the input frequency, giving the relative overall RCA gain--shape, $G_{spr}(\nu)$. The equivalent bandwidth was calculated with
\begin{equation}
\beta_{spr} = {\left(\int{G_{spr}(\nu)d\nu} \right)^2 \over \int{ G_{spr}(\nu)^2 d\nu}}.
\end{equation}

\begin{figure*}
\centering
 \includegraphics[width=\textwidth]{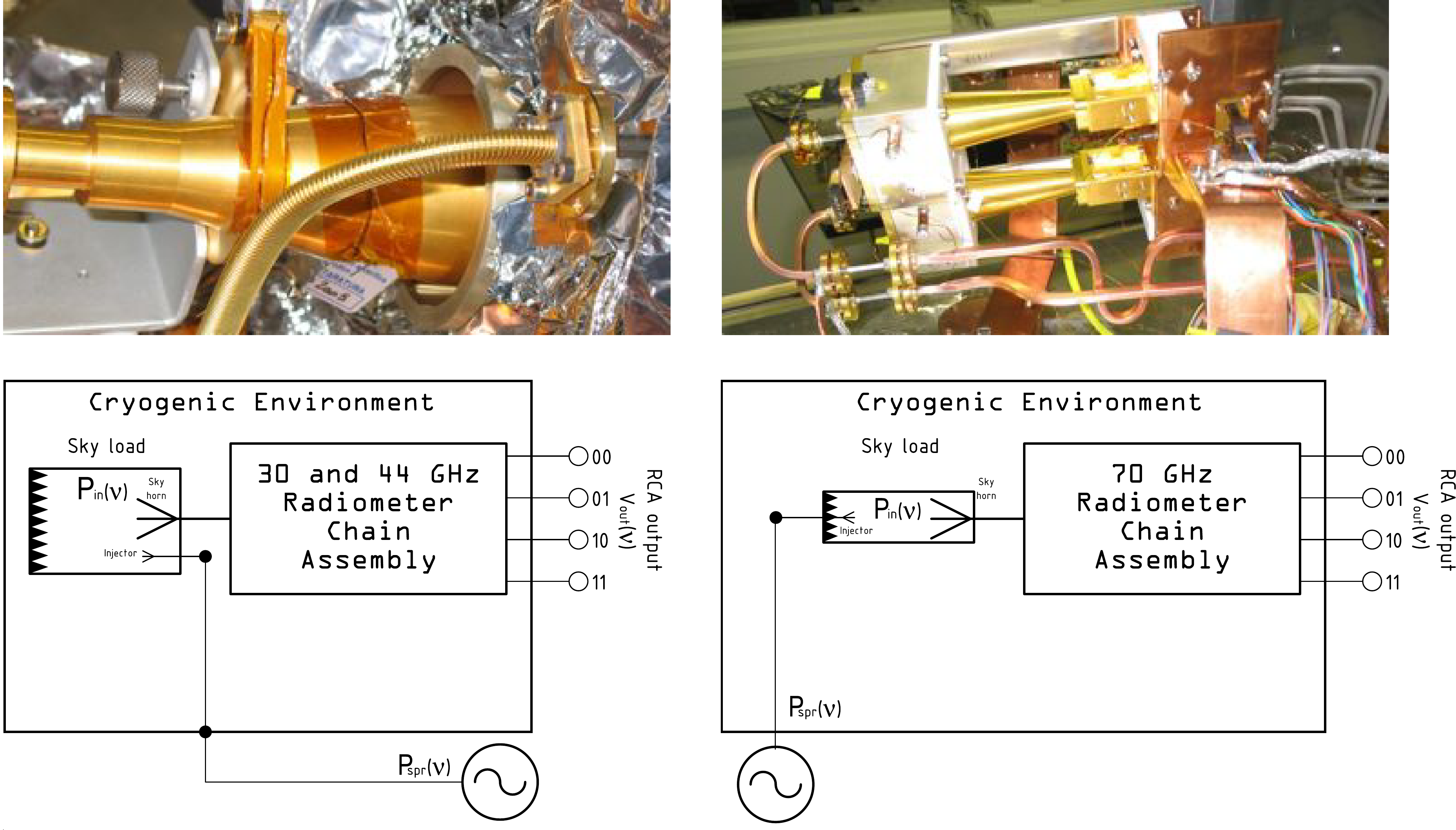} 
   \caption{Setup of the {\tt RCA\_SPR} tests. The picture and the sketch on the left report the test setup of the 30 and 44 GHz RCAs. The flexible waveguide is clearly visible on the picture on the side of the horn. The picture and the sketch on the right report the setup of the 70GHz RCAs. There the signal was injected infront of the horn through the sky load, and copper rigid waveguides were used to carry the signal form the generator to the RCAs.}
   \label{fig:spr_setup}
\end{figure*}

Different setup configurations were used. At 70 GHz the RF signal was directly injected into the sky horn. The input signal was varied by $50$ points from $57.5$GHz to $82.5$GHz.
At 30 GHz and 44 GHz the RF signal was injected into the sky horn after a reflection on the sky load absorber's pyramids, scanning in frequency from $26.5$GHz to $40$GHz in $271$ points and from $33$GHz to $50$GHz in 341 points.  The flexible waveguides WR28 and WR22 were used to reach the skyload for the 30 and 44 GHz RCAs (Fig. \ref{fig:spr_setup}). The input signal was not calibrated in power because only a relative band shape measurement was required. The stability of the signal was ensured by the use of a synthesized sweeper generator guaranteeing the stability of the output within 10\%. The attenuation curve of the waveguide carrying the signal from the sweeper to the injector was treated as a rectangular standard waveguide with losses during the data analysis.

\begin{figure}
\centering
 \includegraphics[width=0.47\textwidth]{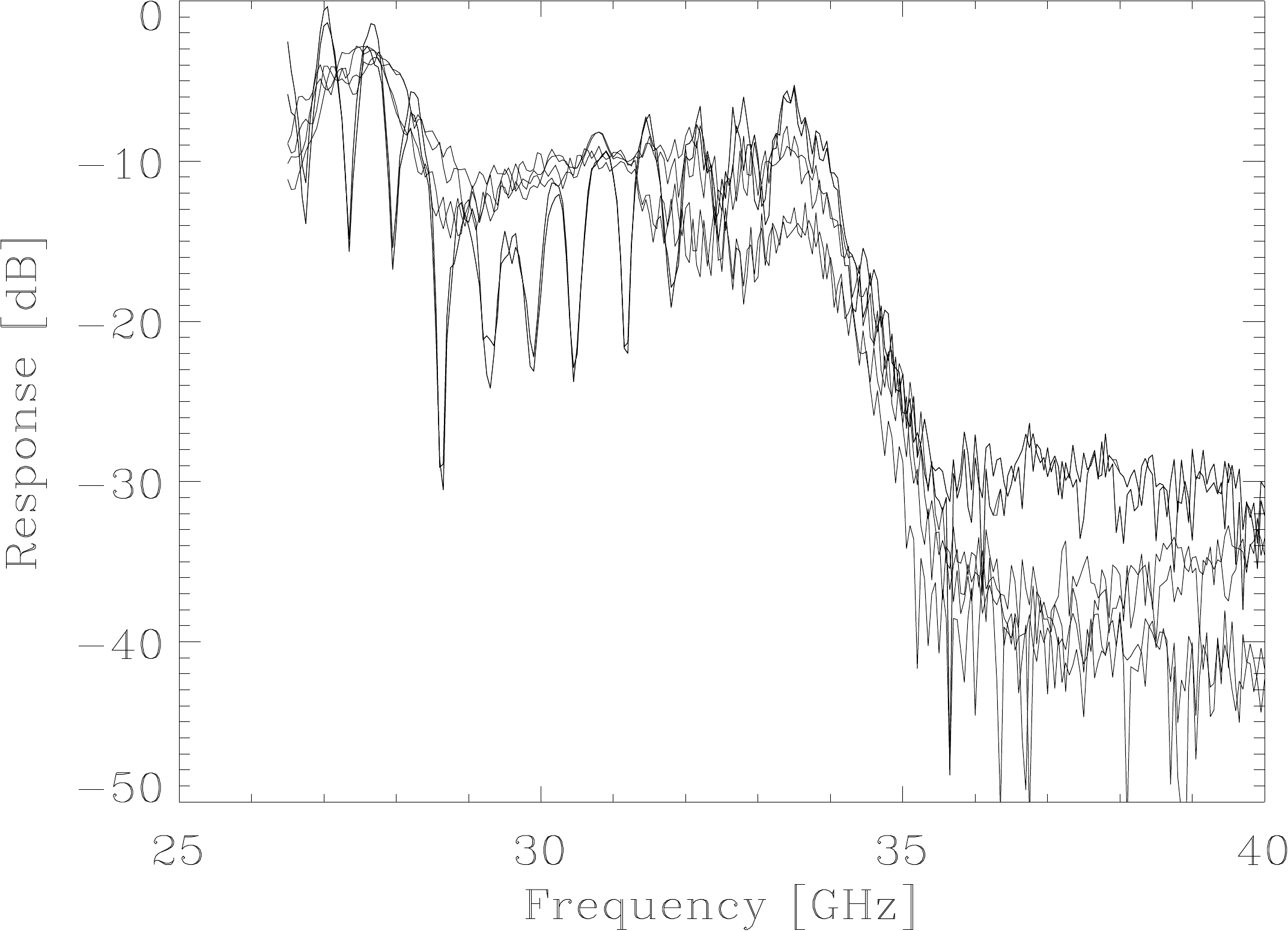} 
   \caption{Measured relative gain function (bandpass) of the 8 detectors at 30 GHz. The curves that show big ripples are those caused by the setup problem (see text). The bandpasses are normalized to the area as explained in the text.}
   \label{fig:spr30}
\end{figure}

\begin{figure}
\centering
 \includegraphics[width=0.47\textwidth]{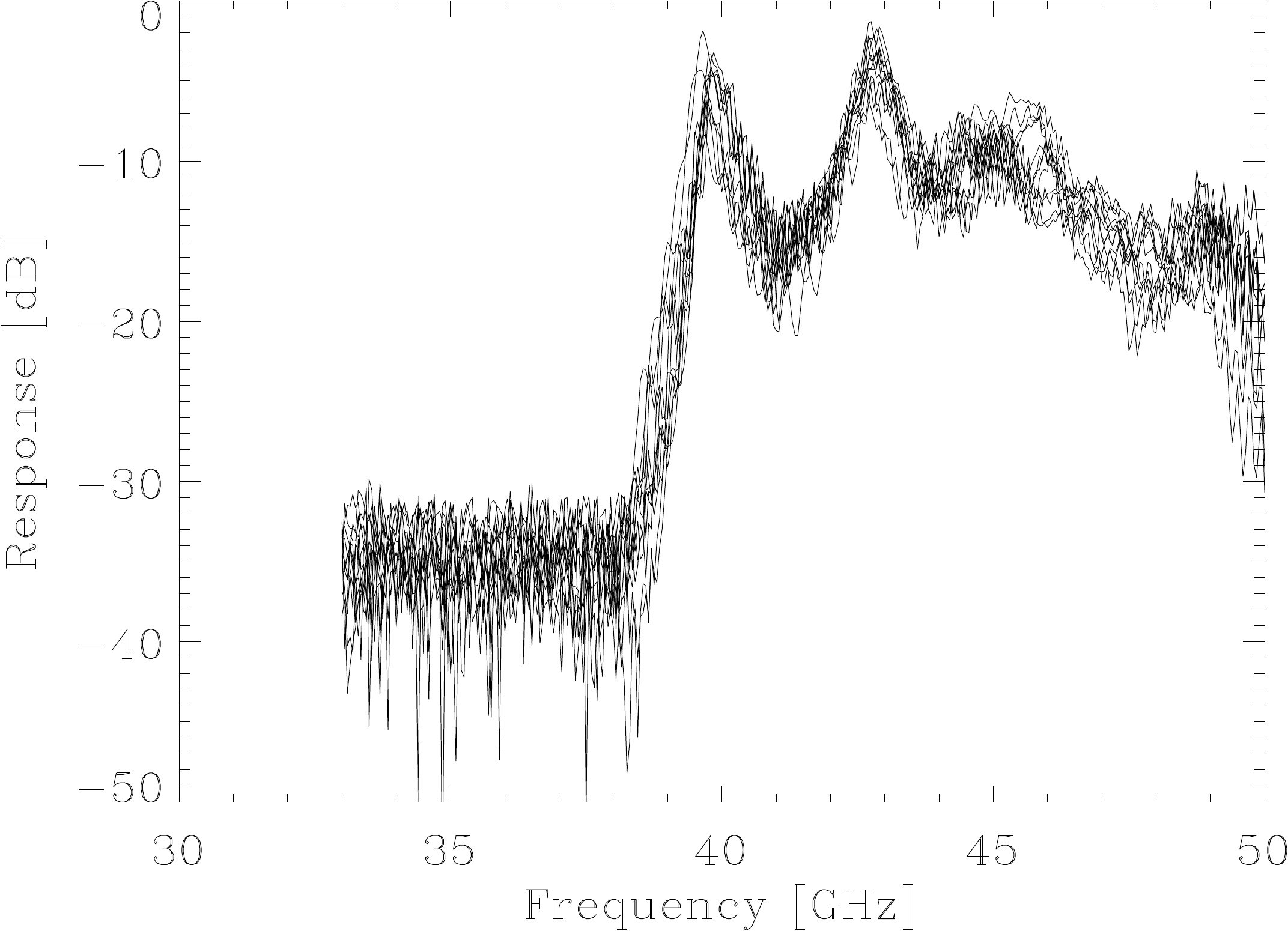} 
   \caption{Measured relative gain function (bandpass) of all 12 detectors at 44 GHz. The bandpasses are normalized to the area as explained in the text.}
   \label{fig:spr44}
\end{figure}

\begin{figure}
\centering
 \includegraphics[width=0.47\textwidth]{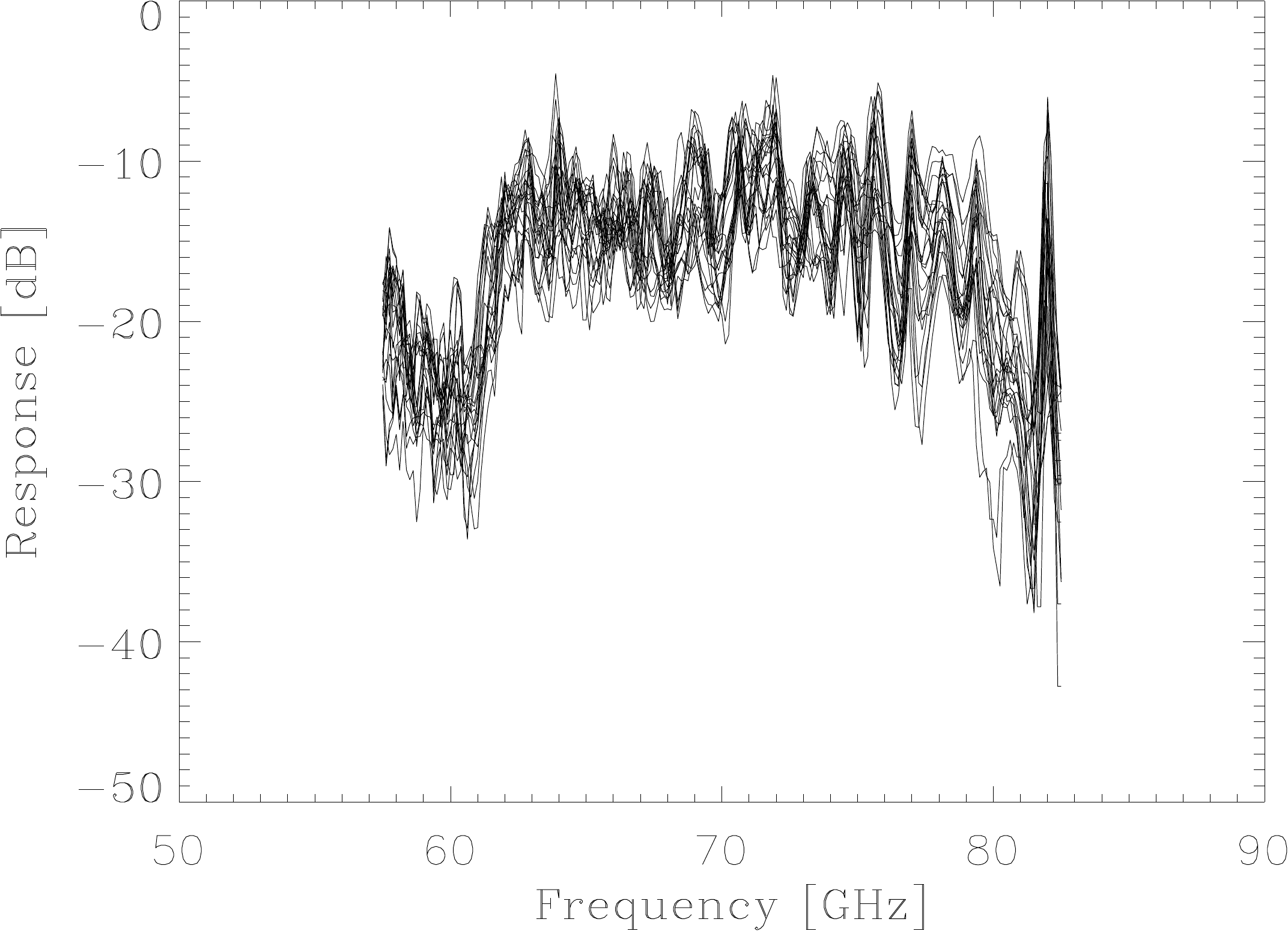} 
   \caption{Measured relative gain function (bandpass) of all 24 detectors at 77GHz. The bandpasses are normalized to the area as explained in the text.}
   \label{fig:spr70}
\end{figure}

All RCA bandshapes were measured, but for the two 30 GHz RCAs only half a radiometer was successfully tested due to a setup problem that appeared when the RCAs were cooled down. For schedule reasons it was not possible to repeat the test at the operational temperature, and only a check at the warm temperature was performed. This warm test was not used for calibration due to different dynamic range, amplifier behavior, and bias conditions.  Results are reported in Table \ref{tab:spr} and plots of all the measurements in Figs. \ref{fig:spr30}, \ref{fig:spr44}, and
\ref{fig:spr70}. All curves reported in the plots are normalized to the area so that 

\begin{equation}
G_{spr}^n(\nu) = {G_{spr}(\nu) \over \int G_{spr}(\nu)d\nu }.
\end{equation}

The bandshape is mainly determined by the filter located inside each BEM, whose frequency response is independent of the tuning of the FEM amplifiers.  The dependence of the bandshape on the amplifier biases has been checked on the 30GHz radiometers  \citep{denardo2008}, showing that at first order the response remains unchanged.  A similar situation occurs on the RCAs at  44 GHz and 70 GHz, where the tuning has second order effects on the overall frequency response.

\begin{table}
\caption{Spectral response test results. The numbers are bandwidth values, $\beta_{spr}$ and central frequency, $\nu_0$, both in GHz.}             
\label{tab:spr}      
\centering                          
\begin{tabular}{l l r r r r}        
\hline\hline                 
  
  & & \multicolumn{4}{c}{\sc{SPR Bandwidth and central frequency }}\\
 & & M-00 & M-01 & S-10 & S-11 \\    
\hline                        
{\tt RCA18}  
& $\beta_{spr}$ &12.40 &11.14	&	10.84&10.25 \\
\hline
{\tt RCA19}  
& $\beta_{spr}$ &10.45 &10.74	&8.00	& 9.91\\
\hline
{\tt RCA20}  
& $\beta_{spr}$ & 11.19&12.21	&	12.57&10.82 \\
\hline
{\tt RCA21}  
& $\beta_{spr}$ & 11.19&	12.21&	12.57& 10.82\\
\hline
{\tt RCA22}  
& $\beta_{spr}$ &11.49 &10.38	&	11.11&10.44 \\
\hline
{\tt RCA23}  
& $\beta_{spr}$ & 10.35&11.52	&	11.62& 11.44\\
\hline
{\tt RCA24}  
& $\beta_{spr}$ &5.15 &	4.08&5.26	&5.82 \\
& $\nu_0$       &45.75 & 42.4& 45.6& 45.3\\
\hline
{\tt RCA25}  
& $\beta_{spr}$ & 4.42&4.49	&4.17	&5.91 \\
& $\nu_0$       &45.75 &45.25&45.85 &44.90 \\
\hline
{\tt RCA26}  
& $\beta_{spr}$ &6.10 &4.86	&4.26	&5.48 \\
& $\nu_0$       &44.35 &44.85 &44.90 & 44.20\\
\hline
{\tt RCA27}  
& $\beta_{spr}$ & --&--	&	3.89& 3.71 \\
& $\nu_0$       & --& --& 30.45& 30.70\\
\hline
{\tt RCA28}  
& $\beta_{spr}$ &4.94 &5.12	&--	& --\\
& $\nu_0$       &31.4& 31.35 &--  & --  \\
\hline
\end{tabular}
\end{table}

\subsection{Susceptibility}
Any variation in physical temperature of the RCA, $T_{phys}$, will produce a variation of the output signal that mimics the variation of the input temperature, $T_{sky}$, so that 

\begin{equation}
\delta T_{sky} = T_f \cdot \delta T_{phys},
\label{eq:thf}
\end{equation}

where $T_{f}$ is the transfer function. A controlled variation of FEM temperature was imposed to calculate the transfer function of the front end modules, $T_{f}^{FEM}$. This was done for all RCAs and all detectors. The chief results are given in Table \ref{tab:thf}, while the details of the applied method and of the measurements are reported by \citet{terenzi2009c}.

\begin{table}
\caption{The transfer function of the susceptibility of FEM to temperature variations.}             
\label{tab:thf}      
\centering                          
\begin{tabular}{l r r r r}        
\hline\hline                 
 & \multicolumn{4}{c}{\sc{FEM susceptibility}}\\
 & \multicolumn{4}{c}{\sc{Transfer Function cK/K}} \\
 & M-00 & M-01 & S-10 & S-11 \\    
\hline                        
{\tt RCA18}  & -8.48 & -9.13 & -6.77 & -7.67 \\
{\tt RCA19}  & -9.43 & -9.28 & -1.20 & -9.49 \\
{\tt RCA20}  & -6.61 & -5.69 & -5.93 & -5.79 \\
{\tt RCA21}  & -3.01 & -1.85 & -0.770 & -0.930 \\
{\tt RCA22}  & 5.67 & 5.21 & 5.69 & 6.04 \\
{\tt RCA23}  & -2.07 & -4.41 & -4.07 & -3.92 \\
{\tt RCA24}  & -1.21  & -0.610 & -2.03  & -0.964 \\
{\tt RCA25}  & -1.51 & -1.33 & -2.87 & -2.21 \\
{\tt RCA26}  & -6.22 & -6.10 & -6.57 & -6.31 \\
{\tt RCA27}  & -1.68 & -1.05 & -3.64 & -3.06 \\
{\tt RCA28}  & -0.266 &-0.519 & -1.85 & -1.05 \\
\hline                        
\end{tabular}
\end{table}

The susceptibility of the radiometer signal to temperature variations in the BEM and 3$^{rd}$ V-groove were measured only for the 30 and 44 GHz chains, because at 70 GHz it was not possible to control the temperatures of these interfaces in their cryofacility.  Here we report on the BEM susceptivity tests only: it is the prominent radiometric effect between both, because the diodes are thermally attached to the BEM body.  The total power output voltage on each BEM detector can then be expressed by modifying Eq. \ref{eq:nl}
\begin{equation}
V_{out} = G_{tot}\left( T_0^{bem}\right)\cdot \left [ 1 +  T_f  \cdot \left ( T^{bem} - T_0^{bem} \right )  \right ] \cdot \left ( T_{in} + T_{sys} \right ),
\end{equation}
There $G_{tot}\left( T_0^{bem}\right)$ is the total power gain when the BEM is at nominal physical temperature, $T_0^{bem}$, while $T^{bem}$ is the BEM physical temperature that was varied in steps. Non-linear effects were not considered because they were mainly caused by the change in $T_{in}$  which was fixed in this case.  By exploiting the linear dependence between  the voltage output and the BEM physical temperature as
\begin{equation}
V_{out} = m\cdot T^{bem} + q,
\end{equation}
with the transfer function, $T_f$, calculated from a linear fit to the data as
\begin{equation}
T_f = {m \over m\cdot T_0^{bem} + q }.
\end{equation}
The values are reported in Table \ref{tab:thb}.

\begin{table}
\caption{The transfer function of the susceptibility of BEM to temperature variations. The physical temperature of the BEM is also reported in the last column.}             
\label{tab:thb}      
\centering                          
\begin{tabular}{l r r r r r}        
\hline\hline                 
 & \multicolumn{4}{c}{\sc{BEM susceptibility}}\\
 & \multicolumn{4}{c}{\sc{Transfer Function 1/K}} \\
 & M-00 & M-01 & S-10 & S-11 & $T_0^{bem}$, $^\circ$C \\    
\hline                        
{\tt RCA24} 	& -0.008	& -0.009	& -0.008	& -0.008	& 34.292 \\
{\tt RCA25	}  &-0.020 &-0.021	&-0.022	&-0.021	&35.237 \\
{\tt RCA26	}  &-0.018	&-0.018	&-0.018	&-0.017	&31.810 \\
{\tt RCA27}	& -0.006	&-0.006	&-0.009	&-0.007	&40.178 \\
\hline                        
\end{tabular}
\end{table}

\section{Conclusions}\label{conclusions}
The eleven LFI Radiometer Chain Assemblies  were calibrated according to the overall LFI calibration plan. The front end low-noise amplifiers and phase switches were properly tuned to guarantee minimum noise temperature and best isolation.  Basic performances (noise temperature, isolation, linearity, photometric gain), noise properties (1/f spectrum, noise equivalent bandwidth), relative bandshape, and susceptibility to thermal variations were measured in a dedicated cryogenic environment as close as possible to flight-operational conditions. 
All radiometric parameters were measured with excellent repeatability and reliability, except for 1/f noise at 70 GHz and some of the bandpasses at 30 GHz. The measurements were essentially in line with the design expectations, indicating a satisfactory instrument performance. Ultimately all
RCA units were accepted, because the measured performances were in line with the scientific goal of LFI. 
The RCA test campaign described here represents the primary calibration test for some key radiometric parameters of the LFI, because they were not accurately measured as part of the RAA calibration campaign, nor can they be measured in-flight during the Planck nominal survey:
\begin{itemize}
\item tuning results were used to set the subsequent tuning verification procedure up to the calibration performance and verification phase (CPV) in flight;
\item non-linear behavior of the 30 and 44 GHz RCAs was accurately measured and characterized and used to estimate the impact in flight \citep{mennella2009b}. Moreover each radiometer system noise temperature was accurately determined; 
\item except for some of the measured spectral responses, in which systematic effects arising from standing waves in the sky load were experienced,  the RCA band shapes were only measured and characterized during the RCA tests. Together with the independent estimates of the band shape based on a synthesis of measured responses at unit level \citep{zonca2009},  they are essential for the flight data analysis; 
\item susceptivity  to thermal variation of the FEM and BEM was measured and represents the reference values because only the FEM susceptivity was measured during RAA tests \citep{terenzi2009b}, but under worse thermal conditions.
\end{itemize}

In conclusion we can state that even if the RCA calibration campaign was an intermediate step in the LFI development, the results obtained and presented here will be used in conjunction with the performance measured in flight to the exploitation of the scientific goal of Planck. 

\begin{acknowledgements}
Planck is a project of the European Space Agency with instruments funded by ESA member states, and with special contributions from Denmark and NASA (USA). The Planck-LFI project is developed by an International Consortium lead by Italy and involving Canada, Finland, Germany, Norway, Spain, Switzerland, UK, USA. 
The Italian contribution to Planck is supported by the Italian Space Agency (ASI). 
In Finland, the Planck LFI 70 GHz work was supported by the Finnish Funding Agency for Technology and Innovation (Tekes)
TP's work was supported in part by the Academy of Finland grants 205800, 214598, 121703, and 121962. TP thank Waldemar von Frenckells stiftelse, Magnus Ehrnrooth Foundation, and V\"ais\"al\"a Foundation for financial support. 
The Spanish participation is funded by Ministerio de Ciencia e Innovacion through the projects
ESP2004-07067-C03-01 and AYA2007-68058-C03-02

\end{acknowledgements}

\bibliographystyle{aa} 
\bibliography{fh} 

\begin{appendix}
\section{Y-factor method}\label{a:Y-factor}

The classical Y-factor method is the simplest way to calculate the noise temperature, and it requires that radiometric data are acquired at two different (possibly well-separated) physical temperatures of one of the input loads. Below we will assume a sky load temperature increase. Clearly the treatment is completely symmetrical if the reference load temperature is increased.
If we denote with $T_{low}$ and $T_{high}$ as the antenna temperatures corresponding to the skyload physical temperatures, we find that the ratio between the output voltages is given by

\begin{equation}
{V_{low} \over V_{high}}  = {T_{low} + T_N \over T_{high} + T_N} \equiv {1\over Y}.
\end{equation}

The system noise temperature is then calculated as 

\begin{equation}
T_N \equiv {T_{high} - Y\cdot T_{low} \over Y-1}.
\end{equation}

\end{appendix}

\begin{appendix}

\section{Radiometer non-linear model (gain model)}\label{app:gain-model}
A non-linear gain model was developed and applied to the 30 and 44 GHz RCAs to model the observed behavior of  the output voltages as a function of input temperature. The model was developed on the basis of \citet{daywitt1989} and specifically adopted for the LFI 30 and 44 GHz RCAs, which exhibit a non-negligible compression effect in the BEMs. The FEM is assumed to have a constant gain and noise temperature 
\begin{equation}
FEM:
\bigg \{
\begin{array}{rl}
Gain = & G_0^{FEM}\\
Noise =& T_{N}^{FEM} \\
\end{array}.
\end{equation}

The BEM \citep{artal2009} shows an overall gain (including the detector diode), which depends on the BEM input power as follows: 

\begin{equation}
BEM:
\bigg \{
\begin{array}{rl}
Gain = & G^{BEM} = {G_0^{BEM} \over 1 + b\cdot G_0^{BEM}\cdot p}\\
Noise =& T_{N}^{BEM} \\
\end{array},
\end{equation}

where $p$ is the power entering the BEM and $b$ is a parameter defining the non-linearity of the BEM. For $b=0$ the radiometer is linear; for $b=\infty$ the BEM has a null-gain for any input power;  for $p=\infty$ the BEM is completely compressed and $G^{BEM}=0$ for any value of the non-linearity parameter. 

The power entering the BEM (we here neglect the attenuation of the waveguides whose effect can be modeled as a small reduction of the FEM gain and a small increase of the FEM noise temperature) can be written as

\begin{equation}
p = k\cdot B\cdot G_0^{FEM} \cdot \left ( T_A + T_N\right ) ,
\end{equation} 
where 
\begin{equation}
T_N  = T_N^{FEM} + {T_N^{BEM}\over G_0^{FEM}}.
 \end{equation}
The voltage at each output BEM detector (the detector diode constant is included in the BEM gain) can be written as 

\begin{eqnarray}\label{g-model}
V_{out} &=& k\cdot B\cdot G_0^{FEM} \cdot 
{G_0^{BEM}\cdot \left (  T_A + T_N\right )  \over 1 + b\cdot G_0^{BEM}\cdot \left (  T_A + T_N\right )} \nonumber \\ 
&=& G_0 \left [ {1\over 1 + b\cdot G_0\cdot \left (  T_A + T_N\right )}\right ] \cdot \left (  T_A + T_N\right ),  \\
\mbox{where} \\
G_0 &=& G_0^{FEM} \cdot G_0^{BEM} \cdot k \cdot B. \nonumber
\end{eqnarray}

In a compact way Eq. (\ref{g-model}) can be written as

\begin{eqnarray}
V_{out} &=& G_{tot} \cdot \left ( T_A + T_N\right ), \\
G_{tot} &=&  G_0 \left [ {1\over 1 + b\cdot G_0\cdot \left (  T_A + T_N\right )}\right ], 
\end{eqnarray}
where the $G_{tot}$ is the total gain of the radiometer, which depends on the input antenna temperature; $G_0$ is the radiometer linear gain and it coincides with the overall gain in case of perfect linearity ($b=0$). 

\end{appendix}

\end{document}